\documentclass[onecolumn, draft, 12pt]{IEEEtran}

\usepackage[utf8]{inputenc}
\usepackage{graphicx}
\graphicspath{{Figs/}}
\usepackage{amssymb,amsmath,amsthm}
\usepackage{subcaption,siunitx,booktabs}
\usepackage[table]{xcolor}
\usepackage{colortbl}
\definecolor{Gray}{gray}{0.85}
\newcolumntype{B}{>{\columncolor{Gray}}c}
\usepackage{psfrag}
\usepackage{pstool}
\usepackage{soul}
\usepackage{diagbox}
\usepackage{cite}


\usepackage{tikz}
\usetikzlibrary{fit,matrix}
\usetikzlibrary{patterns,snakes}

\usepackage{subcaption}
\newtheorem{lemma}{Lemma}
\newtheorem{corollary}{Corollary}
\newtheorem{theorem}{Theorem}
\newtheorem{definition}{Definition}

\newtheorem{remark}{Remark}

\DeclareMathAlphabet{\mathbsf}{OT1}{cmss}{bx}{n}
\DeclareMathAlphabet{\mathssf}{OT1}{cmss}{m}{sl}
\DeclareMathAlphabet{\mathcsf}{OT1}{cmss}{sbc}{n}

\DeclareSymbolFont{bsfletters}{OT1}{cmss}{bx}{n}  
\DeclareSymbolFont{ssfletters}{OT1}{cmss}{m}{n}
\DeclareMathSymbol{\bsfGamma}{0}{bsfletters}{'000}
\DeclareMathSymbol{\ssfGamma}{0}{ssfletters}{'000}
\DeclareMathSymbol{\bsfDelta}{0}{bsfletters}{'001}
\DeclareMathSymbol{\ssfDelta}{0}{ssfletters}{'001}
\DeclareMathSymbol{\bsfTheta}{0}{bsfletters}{'002}
\DeclareMathSymbol{\ssfTheta}{0}{ssfletters}{'002}
\DeclareMathSymbol{\bsfLambda}{0}{bsfletters}{'003}
\DeclareMathSymbol{\ssfLambda}{0}{ssfletters}{'003}
\DeclareMathSymbol{\bsfXi}{0}{bsfletters}{'004}
\DeclareMathSymbol{\ssfXi}{0}{ssfletters}{'004}
\DeclareMathSymbol{\bsfPi}{0}{bsfletters}{'005}
\DeclareMathSymbol{\ssfPi}{0}{ssfletters}{'005}
\DeclareMathSymbol{\bsfSigma}{0}{bsfletters}{'006}
\DeclareMathSymbol{\ssfSigma}{0}{ssfletters}{'006}
\DeclareMathSymbol{\bsfUpsilon}{0}{bsfletters}{'007}
\DeclareMathSymbol{\ssfUpsilon}{0}{ssfletters}{'007}
\DeclareMathSymbol{\bsfPhi}{0}{bsfletters}{'010}
\DeclareMathSymbol{\ssfPhi}{0}{ssfletters}{'010}
\DeclareMathSymbol{\bsfPsi}{0}{bsfletters}{'011}
\DeclareMathSymbol{\ssfPsi}{0}{ssfletters}{'011}
\DeclareMathSymbol{\bsfOmega}{0}{bsfletters}{'012}
\DeclareMathSymbol{\ssfOmega}{0}{ssfletters}{'012}

\newcommand\blfootnote[1]{%
  \begingroup
  \renewcommand\thefootnote{}\footnote{#1}%
  \addtocounter{footnote}{-1}%
  \endgroup
}

\usepackage{lscape}
\usepackage{makecell}
\begin{document}

\allowdisplaybreaks
%

\title{Streaming Erasure Codes over Multi-Access Relayed Networks}
\author{Gustavo Kasper Facenda$^\dagger$, Elad Domanovitz$^\dagger$, Ashish Khisti$^\dagger$, Wai-Tian Tan$^*$ and John Apostolopoulos$^*$ \\ $^\dagger$ University of Toronto, Toronto, ON M5S 3G4, Canada
\\ $^*$ Enterprise Networking Innovation Labs, Cisco Systems, San Jose, CA 95134, USA \blfootnote{Preliminary results for this paper have been published in \cite{Facenda2021MAC}.}  }

\maketitle

\begin{abstract}
	Many emerging multimedia streaming applications involve multiple users communicating under strict latency constraints. In this paper we study streaming codes for a network involving two source nodes, one relay node and a destination node. In our setting, each source node  transmits a stream of messages, through the relay, to a destination, who is required to decode the messages under a strict delay constraint. For the case of a single source node, a class of streaming codes has been proposed by Fong et al. \cite{Silas2019}, using the concept of {\em delay-spectrum}. In the present work we present an in-depth analysis of the properties of delay-spectrum and apply them to develop streaming codes for our proposed setting through a novel framework. Our first scheme involves greedily selecting the rate on the link from relay to destination and using properties of the delay-spectrum to find feasible streaming codes that satisfy the required delay constraints. We provide a closed form expression for the achievable rate region and identify conditions when the proposed scheme is optimal by establishing a natural outer bound. Our second scheme builds upon this approach, but uses a numerical optimization-based approach to improve the achievable rate region over the first scheme. We demonstrate that our proposed schemes achieve significant improvements over baseline schemes based on single-user codes.
\end{abstract}
\begin{IEEEkeywords}
	Cloud Computing, Streaming, Low-Latency, Symbol-Wise Decode-and-Forward, Multi-Access Relay Network, Forward Error Correction, Packet Erasure Channel, Rate Region
\end{IEEEkeywords}


\section{Introduction}\label{intro}
A number of emerging applications including online real-time gaming, real-time video streaming (video conference with multiple users), healthcare (under the name tactile internet), and general augmented reality require efficient low-latency communication. In these applications, data packets are generated at the source in a sequential fashion and must be transmitted to the destination under strict latency constraints. When packets are lost over the network, significant amount of error propagation can occur and suitable methods for error correction are necessary.

There are two main approaches for error correction due to packet losses in communication networks: Automatic repeat request (ARQ) and Forward error correction (FEC). ARQ is {not suitable when considering low latency constraints over long distances, as the round-trip time may be larger than the required delay constraint}. For that reason, FEC schemes are considered more appropriate candidates. The literature has studied codes with strict decoding-delay constraints---called streaming codes---in order to establish fundamental limits of reliable low-latency communication under a variety of packet-loss models. 
Previous works have studied particular, useful cases. In \cite{martinian2004burst}, the authors studied a point-to-point (i.e., two nodes---source and destination) network under a maximal burst erasure pattern. In \cite{leong2012erasure}, the authors have studied, separately, burst erasures and arbitrary erasures. In \cite{badr2013streaming}, the authors have extended the erasure pattern, allowing for both burst erasures and arbitrary erasures. In particular, it was shown that random linear codes \cite{ho2003randomized} are optimal if we are concerned only with correcting arbitrary erasures. Other works that have further studied various aspects of low-latency streaming codes include \cite{JoshiWornell2012, Karzand2017, badr2017layered, badr2017fec, Rashmi2018, krishnan2018rate, fong2019optimal, domanovitz2019explicit, KrishnanLowField2020}.

While most of the prior work on streaming codes has focused on a point-to-point communication link, a network topology that is of practical interest involves a relay node between source and destination, that is, a three-node network. This topology is motivated by numerous applications in which a gateway server, able to decode and encode data, connects two end nodes. Motivated by such considerations, streaming codes for such a setting were first introduced in~\cite{Silas2019}, which derived the time-invariant capacity for the three-node setting, and further extended to a multi-hop network in \cite{domanovitz2020streaming}. A time-variant adaptive code construction has been studied in \cite{AdaptiveRelay} and shown to improve upon the rate of \cite{Silas2019}.

However, a significant part of the mentioned applications, such as real-time gaming and video conferences, involve communications between multiple users and a common server. Motivated by such applications, in this paper we extend the relayed topology of \cite{Silas2019} for a multiple access relay channel. {In \cite{Silas2019}, codes were designed to carefully match the symbol-wise delay from source to relay with the delay from relay to destination, allowing for optimal time-invariant codes. With multiple users, which may have different number of erasures and employ codes with distinct delay patterns, designing codes with such properties is non-trivial. In this paper, we present naive ways of extending codes for the single-user setting to the multi-user setting, and we show that these naive extensions are severely suboptimal, being unable to achieve the sumrate capacity in nontrivial settings. For these reasons, a new framework for designing good codes in the relayed network is required. Such framework is one of our main contributions, as detailed in Section~\ref{sec:maincontr}.}

Similar to the work in \cite{Silas2019}, we focus in time-invariant streaming codes, i.e., the relay node does not change the FEC code as a function of erasures on the source-relay link. This is in contrast to the approach in  \cite{domanovitz2020streaming, AdaptiveRelay}. {This choice is motivated by a variety of reasons, including: no optimal adaptive codes are known for the single-user setting; the known achievable adaptive codes \cite{AdaptiveRelay} for the single-user setting require overheads in order to inform the destination about the erasure pattern from source to relay; in an adaptive code, the delay pattern from source to relay depends on the observed erasure pattern, thus, when generalizing to multiple users, one must take into account numerous possible erasure patterns happening in each link from source to relay, making a general framework hard to design, and analysis extremely complex; for the previous reason, the computational complexity at the relay may also be impractical for such schemes.}
Although we focus on the setting with two source nodes in this paper, it should be noted that the proposed framework and the converse can be directly extended to multiple source nodes, although the expressions become notationally cumbersome.

\subsection{Related Works and Applications}

From a problem formulation perspective, our work is strongly connected to previous works on streaming codes over adversarial channels such as \cite{martinian2004burst, leong2012erasure, badr2013streaming, domanovitz2019explicit, KrishnanLowField2020}. These works use a problem formulation similar to ours, in which an error-free transmission is desired with a strict delay constraint. Works such as \cite{Cohen2019, Cohen2021} study the adaptation of random linear codes in order to provide better performances in dynamic channels, that is, in scenarios where the channel conditions change over time. However, none of these works consider a multi-access setting.

From a setting perspective, many applications benefit from a smaller latency. For example, cloud (or stream) gaming is an application in which the heavy processing of gaming, such as physics computation and rendering, are performed in a cloud server, which then streams the video output to the player, who in turn performs actions which are streamed back to the server. The impact of latency on the user experience and performance on cloud gaming has been widely studied \cite{Jarschel2013, Quax2013, Clincy2013, Claypool2014, Slivar2014, Wen2014, Schmidt2017}. It should be noted that the overall latency is composed by many different types of delay, including local processing, propagation, error correction, server processing and display.

Companies such as Google have put considerable effort in building their own networks in order to minimize propagation delay. Riot Games has done similarly, focused on multi-player online gaming instead of cloud gaming \cite{maynard-koran_2016}. Networks such as WTFast are also used by players in order to reduce their propagation latency and have been evaluated in papers such as \cite{Hains2020}. Here, the delay reduction is achieved by optimized routing, rather than the longer (cheaper) routes that regular Internet Service Providers may allocate \cite{maynard-koran_2016} (often selected to minimize number of hops, as opposed to minimize latency).

Reducing the server-side processing delay has also been studied. For example, Google and Microsoft use machine learning to predict the future inputs from the user \cite{lee2014outatime}, allowing the server processing to happen before it receives the input, achieving what Google calls ``negative latency''. Optimization of the video encoding for cloud gaming has also been considered in the literature \cite{Slivar2015}. Other aspects of latency such as the local processing and display times are mostly influenced by the user's hardware.

On the other hand, the error correcting delay---that is, the amount of time required in order to recover from packet losses---seems to be understudied. As shown in \cite{Carrascosa2020}, the round-trip-delay for Stadia's network is usually between 10 and 15ms. Under a delay constraint of 50ms, re-transmissions taking 10ms represent a significant portion of the delay budget. However, by using FEC such as the ones described in this paper, with packet delays that can be as small as $T = 5$ for considerably lossy channels, lost packets can be recovered within few milliseconds, depending on the packet rate. This is even more significant when considering networks that are not as optimized as Google's, where round-trip delays may be considerably higher.

It should also be mentioned that a setting similar to ours has been studied in \cite{Slivar2019}, where the authors consider a multi-access setting in Cloud Gaming, and show that  Quality-of-Experience (QoE) can be improved by jointly optimizing different streams from different nodes. However, only resource allocation is optimized in the paper, basing the decision on the different types of games analyzed. In our work, we instead focus on optimizing the coding scheme used by the nodes and relay.


\section{System Model and Main Results}

\begin{figure*}
    \centering
    \includegraphics[draft=false,scale=1]{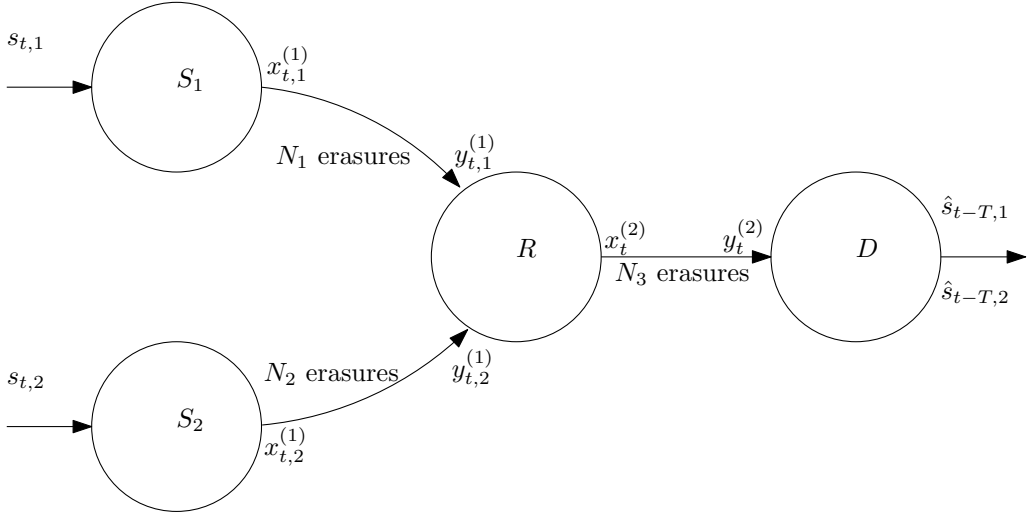}
    \caption{Multiple Access Relay Channel}
    \label{fig:MARC}
\end{figure*}

In this paper, we consider a network with two sources, one relay and one destination. Each source $i$ wishes to transmit a sequence of messages $\{s_{t,i} \}_{t = 0}^{\infty}$ to the destination through a common relay. We assume there is no direct link between sources and destination. We assume that the link between the first source and the relay introduces at most $N_1$ erasures, the link between the second source and the relay introduces at most $N_2$ erasures, and the link between relay and destination introduces at most $N_3$ erasures. The destination wishes to decode both source packets with a common delay $T$. This setting is illustrated in Fig.~\ref{fig:MARC}. This setting captures many practical applications, such as multi-party video conferencing, mentioned previously. Without loss of generality, we assume $N_1 \geq N_2$.

In the following, we present the formal definitions for the problem. For simplicity, we define $\mathbb{F}^n_e = \mathbb{F}^n \cup \{*\}$. The following definitions are standard and a straight-forward generalization of \cite{Silas2019}.
\begin{definition}
    A time-invariant $(n_1, n_2, n_3, k_1, k_2, T, M_1, M_2)_{\mathbb{F}}$-streaming code consists of the following:
    \begin{itemize}
        \item Two sequences of source messages $\{s_{t, 1}\}_{t=0}^{t = \infty}$ and $\{s_{t, 2}\}_{t=0}^{t = \infty}$, where $s_{t, i} \in \mathbb{F}^{k_i}$.
        \item Two encoding functions $$f_{i} : \underbrace{\mathbb{F}^{k_i} \times \cdots \times \mathbb{F}^{k_i} }_{M_i+1\textrm{ times}} \to \mathbb{F}^{n_i}, \quad i \in \{1, 2\}$$
		each used by its respective source $i$ at time $t$ to generate $x^{(1)}_{t, i} = f_{i} (s_{t - M_i, i}, s_{t - M_i + 1, i}, \ldots, s_{t, i})$, where $M_i$ is the memory of the $i$th user encoder.
		\item A relaying function $$g: \underbrace{\mathbb{F}_e^{n_1} \times \cdots \mathbb{F}_e^{n_1}}_{T+1\textrm{ times}} \times  \underbrace{\mathbb{F}_e^{n_2} \times \cdots \mathbb{F}_e^{n_2}}_{T+1\textrm{ times}} \times \underbrace{\mathbb{F}_e^{k_1} \times \cdots \mathbb{F}_e^{k_1}}_{M_1 + 2 - T\textrm{ times}} \times \underbrace{\mathbb{F}_e^{k_2} \times \cdots \mathbb{F}_e^{k_2}}_{M_2 + 2 - T\textrm{ times}} \to \mathbb{F}^{n_3} \times \mathbb{F}^{k_1} \times \mathbb{F}^{k_2} $$
		used by the relay at time $t$ to generate $$(x^{(2)}_t, \hat{s}^{(1)}_{t - T, 1}, \hat{s}^{(1)}_{t - T, 2}) = g( \{y^{(1)}_{j, 1}\}_{j = t - T}^{t}, \{y^{(1)}_{j, 2}\}_{j = t - T}^{t}, \{ \hat{s}^{(1)}_{ \{t - t'\}_{t' = M_1}^{T - 1}}\}, \{ \hat{s}^{(1)}_{ \{t - t'\}_{t' = M_2}^{T - 1}}\} ).$$
		\item Two decoding functions 
		\begin{align*}
		    \varphi_{t,1} = \underbrace{\mathbb{F}_e^{n_3} \times \cdots \times \mathbb{F}_e^{n_3}}_{t+T+1\textrm{ times}} \to \mathbb{F}^{k_1} \\
		    \varphi_{t,2} = \underbrace{\mathbb{F}_e^{n_3} \times \cdots \times \mathbb{F}_e^{n_3}}_{t+T+1\textrm{ times}} \to \mathbb{F}^{k_2} 
		\end{align*}
		used by the destination at time $t + T$ to generate
		\begin{align*}
		    \hat{s}_{t, 1} = \varphi_{t,1}(y^{(2)}_{0},y^{(2)}_{1},\ldots,y^{(2)}_{t+T})\\
		    \hat{s}_{t, 2} = \varphi_{t,2}(y^{(2)}_{0},y^{(2)}_{1},\ldots,y^{(2)}_{t+T})
		\end{align*}
    \end{itemize}
\end{definition}

\begin{remark}
    Note that the function $g$ is fixed, and the only change that occurs from a time instant to another are the input packets. This is to enforce a time-invariant code, and, in particular, a code where the relay does not adapt (change) the FEC code as a function of the erasure pattern that has been observed in the link from sources to relay, for reasons discussed in \ref{intro}. 
\end{remark}

{For the remainder of the paper, we assume $M_1 \to \infty$ and $M_2 \to \infty$, that is, the encoder may depend on all previous packets, and the relay stores all its previous estimates. When considering an infinite-memory code, we denote it simply as a $(n_1, n_2, n_3, k_1, k_2, T)_{\mathbb{F}}$-streaming code.}

\begin{definition}
	An erasure sequence is a binary sequence denoted by $e^{(1)}_{i} \triangleq \{e^{(1)}_{t, i}\}_{t = 0}^{\infty}$, where {$e^{(1)}_{t,i}=\mathbf{1}\{\textrm{an erasure occurs at time $t$ in the link from source $i$ to relay}\}$}. Similarly, $e^{(2)} \triangleq \{e^{(2)}_t\}_{t = 0}^{\infty}$ where $e^{(2)}_t = \mathbf{1}\{\textrm{an erasure occurs at time $t$ in the link from relay to destination} \}$.
	
	An $N$-erasure sequence is an erasure sequence $e$ that satisfies $\sum_{t=0}^{\infty} e_t = N$. In other words, an $N$-erasure sequence specifies $N$ arbitrary erasures on the discrete timeline. The set of $N$-erasure sequences is denoted by $\Omega_N$. 
\end{definition}

\begin{definition} \label{def:channel}
	The mapping $h_n : \mathbb{F}^{n} \times \{0, 1\} \to \mathbb{F}_e^{n}$ of an erasure channel is defined as
	\begin{align}
	h_n(x, e) = \begin{cases}
    x, &\textrm{~if~} e = 0\\
	*, &\textrm{~if~} e = 1  
	\end{cases} \label{eq:erasuremodel}
	\end{align}
	For any erasure sequences $e^{(1)}_{i}$ and any $(n_1, n_2, n_3, k_1, k_2, T)_{\mathbb{F}}$-streaming code, the following input-output relation holds for each $t \in \mathbb{Z}_+$:
	\begin{align}
	y_{t, 1}^{(1)} = h_{n_1}(x^{(1)}_{t, 1}, e^{(1)}_{t, 1})\\
	y_{t, 2}^{(1)} = h_{n_2}(x^{(1)}_{t, 2}, e^{(1)}_{t, 2})\label{eq:erasurerelay}
	\end{align}
	where $e^{(1)}_{i} \in \Omega_{N_i}$, $i \in \{1, 2\}$.
	Similarly, the following input-output relation holds for for each $t \in \mathbb{Z}_+$:
	\begin{align}
	y_{t}^{(2)} = h_{n_3}(x_{t}^{(2)}, e^{(2)}_{t}) \label{eq:erasuredest}
	\end{align}
	where $e^{(2)}_{t} \in \Omega_{N_3}$.
\end{definition}

\begin{definition}
	An $(n_1, n_2, n_3, k_1, k_2, T)_{\mathbb{F}}$-streaming code is said to be $(N_1, N_2, N_3)$-achievable if, for any $e^{(1)}_{t, i}$ and $e^{(2)}_{t}$, for all $t \in \mathbb{Z}_+$ and all $s_{t, i} \in \mathbb{F}^{k_i}$, we have $\hat{s}_{t,i} = s_{t,i}$.
\end{definition}
\begin{definition}
	The pair of rates of an $(n_1, n_2, n_3, k_1, k_2, T)_{\mathbb{F}}$-streaming code is \begin{align*}
	    R_1 &= \frac{k_1}{n}\\
	    R_2 &= \frac{k_2}{n}\\
	    n &= \max(n_1, n_2, n_3)
	\end{align*}
\end{definition}

\begin{definition}
    The capacity (rate) region of an $(N_1, N_2, N_3)$ multi-access relay network under delay constraint $T$ is defined as the set of all rate pairs $(R_1, R_2)$ such that there exists an $(N_1, N_2, N_3)$-achievable $(n_1, n_2, n_3, k_1, k_2, T)_{\mathbb{F}}$-streaming code, where $(R_1, R_2)$ are defined as above.
\end{definition}

\subsection{Main contributions}\label{sec:maincontr}

In this paper, we derive upper and lower bounds to the capacity region of streaming codes in the described setting. Strict delay constraints are relevant in low latency applications, while the adversarial model represents worst-case scenarios, which are commonly of interest in such applications.

The first result we present in the paper is a general upper bound for the capacity region
\begin{theorem}\label{upperbound}
    The capacity region of time-invariant streaming codes for the four-node multi-access relayed network is upper bounded by the following three conditions
    \begin{align}
        R_1 &\leq C(T - N_3, N_1) \\
        R_2 &\leq C(T - N_3, N_2) \\
        R_1 + R_2 &\leq C(T - N_2, N_3)
    \end{align}
    where $C(T, N) = \frac{T + 1 - N}{T + 1}$ is the capacity of a point-to-point channel subject to $N$ erasures and strict delay $T$.
\end{theorem}

Then, for analysis purposes, we consider four different regimes of operation which depend on $N_1$, $N_2$, $N_3$ and $T$. The main focus of study in this paper is the regime we denote as ``weak relay-destination bottleneck''. The term ``relay-destination bottleneck'' denotes the fact that the constraint $R_1 + R_2 \leq C(T - N_2, N_3)$ is active for at least one pair $(R_1, R_2)$, that is, $C(T - N_3, N_1) + (C - N_3, N_2) \geq C(T - N_2, N_3)$, or, in other words, from a sumrate perspective, the relay-to-destination link is the bottleneck. Furthermore, the term ``weak'' refers to such sumrate not being trivially achievable by single-user communication, that is, $C(T - N_2, N_3) \geq C(T - N_3, N_2)$ and $C(T - N_2, N_3) \geq C(T - N_3, N_1)$. For this regime of operation, we present the following result
\begin{theorem} \label{theorem:R2givenR1}
    For any $R_1 \in \left[ 0, C(T - N_3, N_1)\right] $, we have
    \begin{align}
        R_2 \geq \min \left( C(T - N_3, N_2), C(T - N_2, N_3) - C(T - N_3, N_1), R_2'      \right)
    \end{align}
    where
    \begin{align}
        R_2' = \frac{N_2}{N_1} \left[ \frac{N_1}{N_2} - 1 - R_1 + \frac{T + 1 - N_1 - N_3}{T + 1 - N_2} \right].
    \end{align}
\end{theorem}
Theorem~\ref{theorem:R2givenR1} provides a lower bound on the capacity region, however, we note that there is a gap between our lower and upper bounds for certain choices of $N_1$, $N_2$, $N_3$ and $T$. We study the conditions for which the upper bound can be achieved in this scenario and present a sufficient condition for a rate pair to be the capacity. Further, aiming to reduce the gap when this condition is not met, we propose an optimization algorithm for which we present numerical results. This optimization leads to an improved lower bound, albeit still with some gap to the upper bound.

\begin{figure}[]
	\centering
	\includegraphics[draft=false,width=\linewidth]{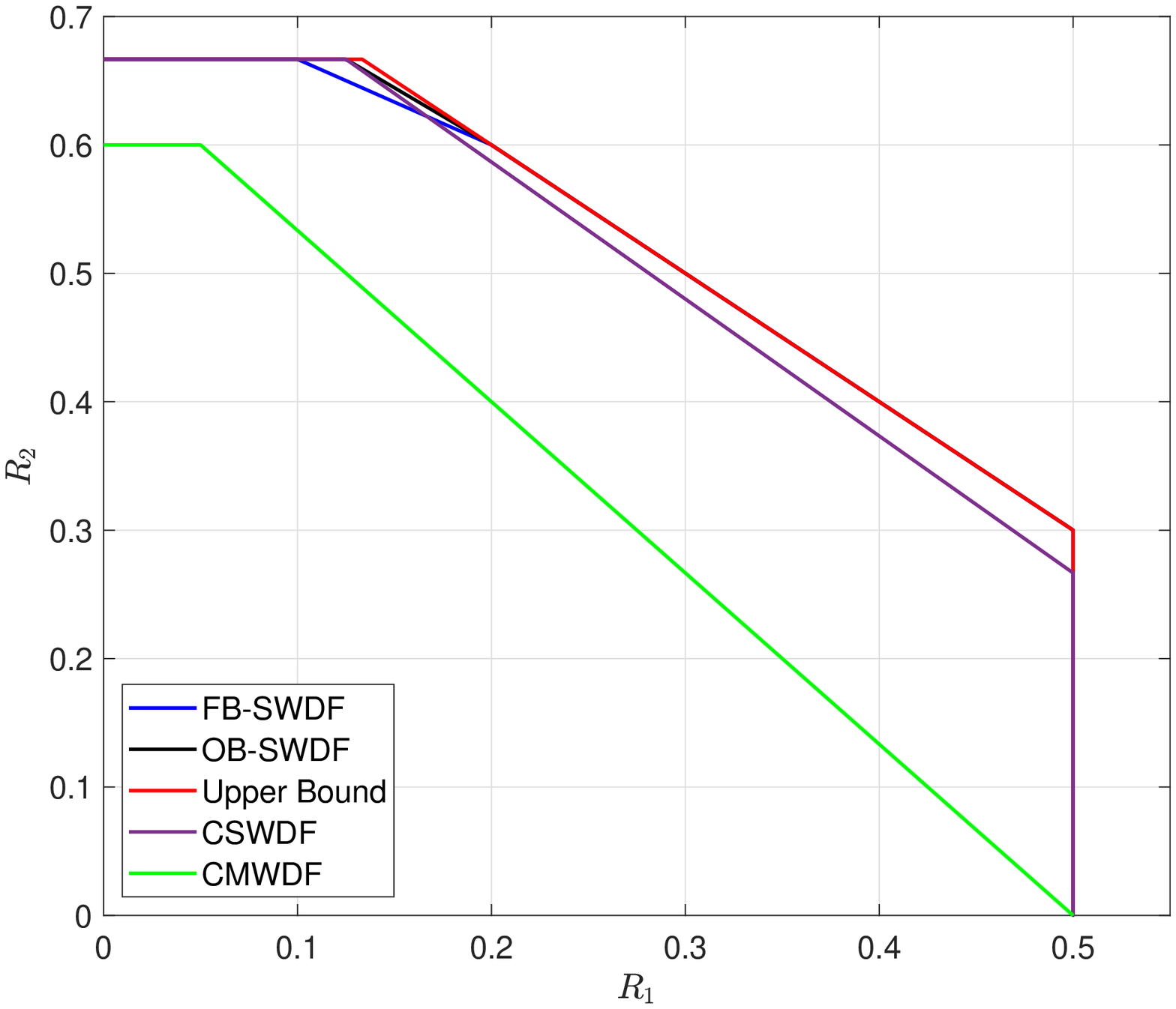}
	\caption{Rate region for $N_1=3,N_2=2,N_3=1,T=6$.}
	\label{fig:introresults}
\end{figure}

For comparison, we also consider naive extensions of schemes known for the three-node relayed network. It can be observed that our optimization outperforms such naive extensions. In particular, it should be noted that these extensions are unable to achieve the sumrate upper bound in any non-trivial scenario. An example of such rate regions can be seen in Fig~\ref{fig:introresults}. In the figure, the rate region derived from Theorem~\ref{theorem:R2givenR1} is denoted as FB-SWDF, the outcome of the optimization method is denoted as OB-SWDF, and the naive extensions of previous methods are CMWDF and CSDWF.

We also present similar lower bound results for the other regimes of operation. In particular, for the so-called ``strong'' regimes, we derive the capacity region, that is, there is no gap between our upper and lower bounds. 

It also should be mentioned that the framework used in this paper in order to derive the lower bounds presented in Theorem~\ref{theorem:R2givenR1} and the optimization algorithm can be easily extended to scenarios with multiple users, to single-user multiplexed coding \cite{Multiplex2018} (in which a user wishes to transmit different streams with different delays to a destination) or to a setting in which each user is subject to different delay constraints. Thus, we believe the contributions made in this paper are significantly broader than the particular setup we study.

\section{Upper bound}\label{sec:upperbound}

In this section, we use the results in \cite{Silas2019} to present a simple upper bound on the achievable rate region. We denote by $C(T, N) = \frac{T + 1 - N}{T + 1}$ the capacity of a single-link point-to-point channel \cite{badr2013streaming}. Since each user is transmitting its own message without cooperation, we have that the following two bounds are direct extensions from \cite{Silas2019}
\begin{align*}
    R_1 &\leq C(T - N_3, N_1)\\
    R_2 &\leq C(T - N_3, N_2).
\end{align*}
{This is because a burst of $N_3$ erasures may occur from time $T - N_3 + 1$ up to $T$, therefore, the relay must have access to all the information about the source packets $s_{t,i}$ at time $T - N_3$, otherwise, it will not be able to relay the remaining information. Thus, we can bound each link from source $i$ to relay as a point-to-point link with effective delay $T - N_3$ and $N_i$ erasures, obtaining the bound above.}

Furthermore, we can optimistically consider $N_2$ erasures in both links in the first hop, and obtain the following upper bound on the sumrate
\begin{align*}
    R_1 + R_2 \leq C(T - N_2, N_3).
\end{align*}
{This follows from a similar argument. Because the relay is required to be time-invariant\footnote{If the relaying function is not required to be time-invariant, higher rates can be achieved, at the cost of overhead and an adaptive relaying scheme that depends on the erasure pattern observed in each link.}, the FEC code employed by the relay may assume that it will only recover any information about the source packets, at best, at time $t + N_2$, since a burst of $N_2$ erasures may happen in both links from sources to relay. Then, the link from relay to destination can be bounded by a point-to-point link with effective delay $T - N_2$ that must handle $N_3$ erasures.}

\section{Symbol-wise decode and forward and delay spectrum}\label{sec:p2pdf}

In order to present our coding scheme, first let us define the notion of delay spectrum for a point-to-point, single-link code. This notion exists and is mentioned in \cite{Silas2019}, however, we make an in-depth analysis of this concept, which has not been made before. Similar analysis has been done in works such as \cite{Multiplex2018}, where a source wishes to transmit two streams with different delays to a destination, however, we generalize it for any number of different streams and delays, focusing on the arbitrary erasure channel instead of the burst channel.
\begin{definition}
	An $(n, k, \mathbf{T}, M)_{\mathbb{F}}$ point-to-point code, where $\mathbf{T} = [T[1], \ldots, T[k]]$ is the delay spectrum of the code, consists of the following:
	\begin{enumerate}
		\item A sequence of source messages $\{s_t\}_{t = 0}^{\infty}$, where $s_t \in \mathbb{F}^{k}$.
		\item An encoding function
		\begin{align*}
		f &: \underbrace{\mathbb{F}^{k} \times \cdots \times \mathbb{F}^{k} }_{M+1\textrm{ times}} \to \mathbb{F}^{n}
		\end{align*}
		used by the transmitter at time $t$ to generate $$x_{t} = f (s_{t - M}, s_{t - M + 1}, \ldots, s_t).$$
		\item A list of $k$ decoding functions 
		\begin{align*}
		\varphi_{t+T[j]} &= \underbrace{\mathbb{F}_e^{n} \times \cdots \mathbb{F}_e^{n}}_{t+T[j]+1\textrm{ times}} \to \mathbb{F}^{k}
		\end{align*}
		where $j \in \{1, 2, \ldots, k\}$ is the index of the $j$th symbol (element of $s_t$), used by the receiver at time $t + T[j]$ to generate $\hat{s}_t[j]$, that is, an estimation of the $j$th element of $s_t$.
	\end{enumerate}
\end{definition}

Similar to before, we assume $M \to \infty$ and refer only to $(n, k, \mathbf{T})_{\mathbb{F}}$ point-to-point streaming codes. 

\begin{definition}
	An $(n, k, \mathbf{T})_{\mathbb{F}}$ point-to-point code is said to achieve delay spectrum $\mathbf{T}$ under $N$ erasures if, for any $e_{t}' \in \Omega_{N}$, 
	\begin{align*}
	\varphi_{t + T[j]}( h_n(x_{0}, e_{0}'), \ldots, h_n(x_{t + T[j]}, e'_{t + T[j]})) = s_t[j]
	\end{align*}
\end{definition}

For the relaying strategy, let us now introduce the concept of symbol-wise decode-and-forward. In this strategy, the relaying function employed by the code first decodes the source packets transmitted by the source, and then encodes them again. This is an extension of the symbol-wise decode-and-forward defined in \cite{Silas2019} for the three-node network. However, the addition of a second source node adds some nuances to the strategy, as the messages relayed by the relay now must be multiplexed in some way. Below, we formally define this strategy.

For the remaining of the paper, we use $\mathbf{T}_i^{(1)}$ to denote the delay spectrum of the code used by the source node $i$ and $\mathbf{T}^{(2)}$ to denote the delay spectrum of the code employed by the relay. 

\begin{definition}\label{def:swdf}
	Assume the source nodes transmit their source messages $\{s_{t,i}\}_t^{\infty}$ to the relay using an $(n_i, k_i, \mathbf{T}_i^{(1)})_{\mathbb{F}}$ point-to-point, single-link code with decoding functions $\varphi^{(1)}_{t+ T_i^{(1)}[j], i}$. Then, a relay is said to employ a symbol-wise decode-and-forward if the following holds:
	\begin{itemize}
		\item The relay employs the decoding functions $\varphi^{(1)}_{t + T_i^{(1)}[j], i}$ at time $t + T_i^{(1)}[j]$ to estimate the $j$-th source symbol of each source packet $\hat{s}_{t,i}[j]$.
		\item The relay employs an $(n_3, k_1 + k_2, \mathbf{T}^{(2)})_{\mathbb{F}}$ point-to-point, single-link code to transmit a sequence of relay messages $\{s'_t \}_{t = 0}^{\infty}$ to the destination.
		\item The relay messages are given by
 		\begin{align}
 		    s'_t[j'] = \hat{s}_{t - (T^{(1)}[j]), 1}[j_1]
 		\end{align}
 		where $j_1 \in \{1, 2, \ldots, k_1\}$, or
 		\begin{align}
 		    s'_t[j'] = \hat{s}_{t - (T^{(1)}[j]), 2}[j_2]
 		\end{align}
 		where $j_2 \in \{1, 2, \ldots, k_2\}$.
	\end{itemize}
\end{definition}

This definition simply implies that each source employs a point-to-point code, the relay attempts to decode the symbols from each source, then re-encodes them together using another point-to-point code. Before we continue, let us give a brief and simple example of a symbol-wise decode-and-forward code designed for channel parameters $N_1 = 1$, $N_2 = 1$, $N_3 = 1$ and $T = 2$. Both source nodes employ the point-to-point link presented in Table~\ref{tab:example1}. Then, the relay decodes every symbol with a delay of 1, that is, each symbol $s_{t, i}[1]$ is recovered at time $t + 1$. The relay then employs the following rule: $s'_t[1] = s_{t - 1, 1}[1]$ and $s'_t[2] = s_{t - 1, 2}[1]$. That is, the relay will now encode using a point-to-point code with $k = 2$. The first symbol of this code will be a delayed version of the source symbol coming from the first user, while the second symbol of this code will be a delayed version of the source symbol coming from the second user. This code is shown in Table~\ref{tab:example2}. {In this example, we have $k_1 = 1$, $k_2 = 1$, $n_1 = n_2 = 2$ and $n_3 = 4$. Thus, we have $R_1 = R_2 = 1/4$. This matches the upper bound $R_1 + R_2 \leq 1/2$. From the table, it can be seen that, even with any one erasure from sources to relay, the relay is still able to recover the desired symbols with a delay of 1 time instant. Similarly, the destination is able to recover any symbol with delay at most $T = 2$ with any one erasure from relay to destination.}

\begin{table}[]
\centering
\caption{Example of symbol-wise decode-and-forward for $N_1 = 1$, $N_2 = 1$, $N_3 = 1$, $T = 2$}
\begin{subtable}{\textwidth}
\centering
\caption{Code used from source $i$ to relay} \label{tab:example1}
\begin{tabular}{|c|c|c|c|c|c|c|c|c|}
\hline
0          & 1          & 2          & 3          & 4          & 5          & 6          & 7          & 8          \\ \hline
$s_{0, i}[1]$ & $s_{1, i}[1]$ & $s_{2, i}[1]$ & $s_{3, i}[1]$ & $s_{4, i}[1]$ & $s_{5, i}[1]$ & $s_{6, i}[1]$ & $s_{7, i}[1]$ & $s_{8, i}[1]$ \\ \hline
           & $s_{0, i}[1]$ & $s_{1, i}[1]$ & $s_{2, i}[1]$ & $s_{3, i}[1]$ & $s_{4, i}[1]$ & $s_{5, i}[1]$ & $s_{6, i}[1]$ & $s_{7, i}[1]$ \\ \hline
\end{tabular}
\end{subtable}

\begin{subtable}{\textwidth}
\centering
\caption{Code used from relay to destination} \label{tab:example2}
\begin{tabular}{|c|c|c|c|c|c|c|c|}
\hline
0 & 1                       & 2                       & 3                       & 4                       & 5                       & 6                       & 7                       \\ \hline
  & $s'_1[1] = s_{0, 1}[1]$ & $s'_2[1] = s_{1, 1}[1]$ & $s'_3[1] = s_{2, 1}[1]$ & $s'_4[1] = s_{3, 1}[1]$ & $s'_5[1] = s_{4, 1}[1]$ & $s'_6[1] = s_{5, 1}[1]$ & $s'_7[1] = s_{6, 1}[1]$ \\ \hline
  &                         & $s'_1[1]$               & $s'_2[1]$               & $s'_3[1]$               & $s'_4[1]$               & $s'_5[1]$               & $s'_6[1]$               \\ \hline
  & $s'_1[2] = s_{0, 2}[1]$ & $s'_2[2] = s_{1, 2}[1]$ & $s'_3[2] = s_{2, 2}[1]$ & $s'_4[2] = s_{3, 2}[1]$ & $s'_5[2] = s_{4, 2}[1]$ & $s'_6[2] = s_{5, 2}[1]$ & $s'_7[2] = s_{6, 2}[1]$ \\ \hline
  &                         & $s'_1[2]$               & $s'_2[2]$               & $s'_3[2]$               & $s'_4[2]$               & $s'_5[2]$               & $s'_6[2]$               \\ \hline
\end{tabular}
\end{subtable}
\end{table}

Furthermore, we define a concatenation of point-to-point codes. 
\begin{definition}
	A concatenation of an $(n', k', \mathbf{T}')_{\mathbb{F}}$ point-to-point code with an $(n'', k'', \mathbf{T}'')_{\mathbb{F}}$ point-to-point code is an $(n' + n'', k' + k'', [\mathbf{T}', \mathbf{T}''])$ point-to-point code with the following properties
	\begin{itemize}
		\item Let $\{f'_{t}\}$ be the encoding function of the first code and $\{f''_{t}\}$ be the encoding function of the second code. The encoding function of the concatenated code is given by $\{[f'_{t}, f''_{t}]\}$, where $[x, y]$ denotes the concatenation of a vector $x$ and a vector $y$.
		\item Let $\{\varphi'_{t + T'[j]}\}_{j=1}^{k'}$ be the list of decoding functions of the first code and $\{\varphi''_{t + T''[j]}\}_{j=1}^{k''}$ be the list of decoding functions of the second code. The list of decoding functions of the concatenated code is given by $\{\varphi'_{t + T'[j]}\}_{j=1}^{k'} \frown \{\varphi''_{t + T''[j]}\}_{j=1}^{k''}$, where $x \frown y$ denotes the concatenation of two lists $x$ and $y$.
	\end{itemize}
	\label{def:concatenation}
\end{definition}

{In the previous example, in Table~\ref{tab:example2}, one can see such concatenation. Note that the code used from relay to destination is a concatenation of two diagonally-interleaved MDS codes with $n' = 2$ and $k' = 1$. This results in a code with $n = 4$ and $k = 2$.}

\begin{lemma}\label{lemma:conc}
	If there exists an $(n', k', \mathbf{T}')_{\mathbb{F}}$ point-to-point code that achieves delay spectrum $\mathbf{T}'$ under $N$ erasures, and an $(n'', k'', \mathbf{T}'')_{\mathbb{F}}$ point-to-point code that achieves delay spectrum $\mathbf{T}''$ under $N$ erasures, then there exists an $(n' + n'', k' + k'', [\mathbf{T}', \mathbf{T}''])_{\mathbb{F}}$ point-to-point code that achieves delay spectrum $[\mathbf{T}', \mathbf{T}'']$ under $N$ erasures.  
\end{lemma}

Another useful operation that can be made is simply permuting the source symbols.
\begin{lemma}\label{lemma:permutation}
	Assume a delay spectrum $\mathbf{T}$ is achievable under $N$ erasures by some code. Then, any permutation $\pi \mathbf{T}$, where $\pi$ is a permutation matrix, of this delay spectrum is also achievable under $N$ erasures.
\end{lemma}

Since any permutation of an achievable delay spectrum is also achievable, we may instead describe the delay spectrum of a code by stating how many symbols are transmitted with some delay.

\begin{definition}
	Consider a delay spectrum $\mathbf{T} = [T[1], T[2], \ldots, T[k]]$. An equally-delayed-symbols grouping description of such delay spectrum is given by a list of tuples
	$$\mathbf{G} = [(T^{(g)}[1], k^{(g)}[1]),\ldots,(T^{(g)}[\ell^{(g)}], k^{(g)}[\ell^{(g)}])]$$ where $\ell^{(g)}$ is the length of the list. For simplicity, we assume $T^{(g)}[1] \geq T^{(g)}[2] \geq \cdots \geq T^{(g)}[\ell^{(g)}]$, therefore, $T^{(g)}[1] = \max(\mathbf{T})$ and $T^{(g)}[\ell^{(g)}] = \min(\mathbf{T})$.  Furthermore, we define
	\begin{itemize}
		\item $\mathbf{T}^{(g)} = [T^{(g)}[1], \ldots, T^{(g)}[\ell^{(g)}]]$ as the ordered list of possible delays.  
		\item $\mathbf{k}^{(g)} = [k^{(g)}[1], \ldots, k^{(g)}[\ell^{(g)}]]$, where $\sum_{i = 1}^{\ell^{(g)}} k^{(g)}[i] = k$, as the ordered list of number of symbols associated with each delay. 
	\end{itemize}
\end{definition}

{Again, referring to the previous example, we could describe the code used in Table~\ref{tab:example2} with $\mathbf{G} = [(1, 2)]$, that is, two symbols are recovered with delay 1.} 

\subsection{Single-link point-to-point results for delay spectrum}

In this Section, we present an achievability result for single-link point-to-point codes in terms of delay spectrum.
\begin{lemma}[Achievability] \label{lemma:achiev}
	Let \begin{equation*}
	\mathbf{T}^{(g)} = \left[T^{(g)}[1], T^{(g)}[2], \ldots, T^{(g)}[\ell^{(g)}]] = [T^{(g)}[1], T^{(g)}[1] - 1, \ldots, N+1, N \right]
	\end{equation*}
	and $\frac{n-k}{N}$ be an integer.
	Then, there exists an $(n, k, \mathbf{T})_{\mathbb{F}}$ point-to-point single-link code that can transmit $k^{(g)}[1] = n - \frac{T^{(g)}[1]}{N}(n-k)$ symbols with delay $T^{(g)}[1]$ and $k^{(g)}[j] = \frac{n - k}{N}$ for all other delays.
\end{lemma}

The above Lemma can be used to derive the following condition on achievability:
\begin{corollary}\label{corollary:maxsym}
	Assume there is a maximum number of symbols we are allowed to transmit at each delay $T^{(g)}[j]$, that is, $\mathbf{k}^{(g)} \leq \mathbf{k}^{\textrm{con}}$ is a constraint. Then, if
	\begin{align}
	k &\leq n - n \cdot N \cdot \frac{\left( 1 - \sum_{\ell = 1}^{j-1} \frac{k^{\textrm{con}}[\ell]}{n}  \right)}{T^{(g)}[j] + 1} \forall j \in \{1, 2, \ldots, \ell^{(g)}\} \\
	k &\leq \sum_{\ell = 1}^{\ell^{(g)}} k^{\textrm{con}}[\ell]
	\end{align}
	there exists an $(n, k, \mathbf{T})_{\mathbb{F}}$ point-to-point single-link code that achieves the desired delay spectrum $\mathbf{T}^{(g)}$ under $N$ erasures.
\end{corollary}

The usefulness of this corollary should be clearer when we present our Fixed-Bottleneck Symbol-wise Decode-and-forward scheme.

\section{Achievable rate region} \label{sec:achievable}

In this section, we present lower bounds on the capacity region. For now, let us make the following assumptions on the parameters:
\begin{itemize}
    \item $N_1 \geq N_2 \geq N_3$. Recall that we had already assumed $N_1 \geq N_2$ without loss of generality.
    \item $T \geq \frac{1}{2}\left( \sqrt{N_1^2 - 4N_3(N_2 - N_3)} + N_1 + 2N_2 - 2 \right)$.
\end{itemize}
These conditions ensure that we are operating in the so-called ``weak relay-destination bottleneck'' regime. Later, in Section~\ref{sec:allregimes}, we study all the regimes of operation.

We start the section by introducing a novel scheme, in which the relay-destination link attempts to transmit at its maximal rate, denoted as Fixed-Bottleneck Symbol-Wise Decode-and-Forward (FB-SWDF). This scheme provides us with the lower bound in Theorem~\ref{theorem:R2givenR1}. Then, we formally present tools that allow us to naively extend the known schemes for the three-node network to a multi-access network. In particular, one of these schemes---denoted Concatenated Symbol-Wise Decode-and-Forward (CSWDF)---is able to outperform our FB-SWDF scheme for some rate pairs. This is due to the attempt of transmitting at maximal rate. Finally, we present an optimization approach, which allows us to achieve the highest known rates. 

Examples of the rates achieved by each scheme and the code construction of our FB-SWDF are presented in Section~\ref{sec:example}.

\subsection{Fixed-Bottleneck Symbol-Wise Decode-and-Forward}\label{sec:fixedrate}

To give a general idea of the scheme, let us first attempt to fix the code employed by the relay. Let us use a code with rate $R_{bn} = C(T - N_2, N_3)$, where $R_{bn}$ is the rate of the single-link code employed by the relay. For some choice of $n_3$, we then can find the delay spectrum of this code by employing Lemma~\ref{lemma:achiev}. Note that the delay spectrum of the code employed by the relay imposes a constraint on the delay spectrum of the codes employed by the source nodes.

Further, let us also fix some code for the first user. In particular, let us attempt to transmit with rate $R_1 = C(T - N_3, N_1)$, i.e., its single-user capacity. Then again, for some choice of $n_1$, using Lemma~\ref{lemma:achiev}, we are able to find the delay spectrum of this code. This will, in turn, update the imposed constraints on the delay spectrum of the second user.

Finally, with the updated delay constraints for the second user, we can apply Corollary~\ref{corollary:maxsym} and find an achievable rate for $R_2$ given the delay constraints imposed by the codes used in the relay and from the first source node to relay.

With the correct choice of $n_1$, $n_2$ and $n_3$, this coding scheme allows us to derive the following Lemma.
\begin{lemma}\label{lemma:achievablelowercorner}
	For $R_1 = C(T - N_3, N_1)$ the following rate is achievable in the second link
	\begin{align}
		R_2 = \min(R_2', C(T - N_2, N_3) - C(T - N_3, N_1))
	\end{align}
	where
	\begin{align}
		R_2' = 1 - \frac{N_2 (1 - (T + 1 - N_3 - N_1)\frac{(N_2 - N_3)}{(T + 1 - N_2)(T + 1 - N_3)}) }{N_1}
	\end{align}
\end{lemma}

One can notice that this Lemma is a particular case of Theorem~\ref{theorem:R2givenR1}. However, we believe that introducing this coding scheme first, and then generalizing it, allows a clearer understanding of the steps required in order to achieve the rates presented. Furthermore, this Lemma is useful because it allows us to derive the following corollary, which gives a sufficient condition for the sumrate to be achieved for at least one rate pair. 

\begin{corollary}\label{corollary:minimumT}
	If $(T + 1 - N_3)(T + 1 - N_2 - N_1) \geq (T + 1 - N_3 - N_1)(N_2 - N_3)$, then the sumrate capacity $R_1 + R_2 = C(T - N_2, N_3)$ is achieved at at least one point of the rate region.
\end{corollary}

Now, note that we can follow the steps mentioned in this section for any $R_{bn}$ and $R_1$. In fact, Theorem~\ref{theorem:R2givenR1} is obtained by simply considering a general $R_1$, instead of $R_1 = C(T - N_3, N_1)$. In the derivation of Theorem~\ref{theorem:R2givenR1}, we actually consider a code that might seem suboptimal for the first source node---we employ a slightly deteriorated delay spectrum, in order to obtain a uniform delay spectrum. For example, instead of transmitting 2 symbols with delay 1 and zero symbols with delay 2, we may opt to transmit 1 symbol with each delay instead. Numerical results show that this choice of code does not, in fact, decrease the rate achieved by the second user. That is, the achievable rate region obtained by employing such codes in the link from first user to relay or by employing an optimal-delay-spectrum code is the same. Below, we restate Theorem~\ref{theorem:R2givenR1}.
\newtheorem*{theorem*}{Theorem}
\begin{theorem*}
	For any $R_1$, the following rate is achievable in the second link in the first hop
	\begin{equation}
		R_2 = \min \left( C(T - N_3, N_2), C(T - N_2, N_3) - C(T - N_3, N_1), R_2' \right)
	\end{equation}
	where
	\begin{equation}
		R_2' = \frac{N_2}{N_1}\left[ \frac{N_1}{N_2} - 1 - R_1 + \frac{T + 1 - N_1 - N_3}{T + 1 - N_2} \right]
	\end{equation}
\end{theorem*}

Furthermore, we are also interested in knowing which part of the capacity region (in particular, in the sumrate region) is achievable, when some of it is. This is provided by the following Lemma, which provides another achievable rate pair. This rate pair is simply the intersection between the line defined by $R_2'$ in Theorem~\ref{theorem:R2givenR1} and the sumrate bound. 

\begin{lemma} \label{lemma:intersectionpoint}
	If the sumrate capacity is achieved at at least one point of the rate region, then the following point is also achievable
	\begin{align}
		R_1 = \frac{N_2 - N_3}{T + 1 - N_2}\\
		R_2 = \frac{T + 1 - 2N_2}{T + 1 - N_2}
	\end{align}
\end{lemma}

The complete proofs of all the statements in this section are presented in the Appendix, with explicit description of the code parameters used.

\subsection{Extensions of Known Schemes}

In this section, we present a time-sharing-like tool that allows us to generalize the results in \cite{Silas2019}, and can also be used to define achievable regions based on achievable points. In particular, using this lemma in combination with Lemma~\ref{lemma:achievablelowercorner} and Lemma~\ref{lemma:intersectionpoint} allows us fully characterize the part of the region in which we achieve the capacity.

Using the following Lemma, we can extend the schemes presented in \cite{Silas2019} and, generally, show that if any two points are achievable, (practically) any linear combination between such two points is also achievable.
\begin{lemma}\label{lemma:timesharing}
	The following lemma holds for symbol-wise decode-and-forward: assume there exists an $(N_1, N_2, N_3)$-achievable $(n_1, n_2, n_3, k_1, k_2, T)_{\mathbb{F}}$-streaming code, and another $(N_1, N_2, N_3)$-achievable $(n'_1, n'_2, n'_3, k'_1, k'_2, T)_{\mathbb{F}}$-streaming code. Then, for any $A, B \in \mathbb{Z}$, there exists an $(N_1, N_2, N_3)$-achievable $(A n_1 + B n_1', An_2 + B n_2', A n_3 + B n_3', A k_1 + B k_1', A k_2 + B k_2', T)_{\mathbb{F}}$-streaming code.
\end{lemma}
\begin{remark}
    While Lemma~\ref{lemma:timesharing} allows us to achieve any convex combination of two pairs of rates $(R_1, R_2)$ and $(R_1', R_2')$, it should be noted that it is possible that, due to the nature of the definition of $n$ in the rate, we achieve better than the simple convex combination. This can be seen in Section~\ref{sec:results} for both the CSWDF and CMWDF schemes, where the achievable rate region of such schemes is not a straight line connecting the two single-user rates.
\end{remark}

\subsubsection{Concatenated Symbol-wise Decode-and-forward} \label{sec:concSW}

The first scheme we consider is an extension of the single-user capacity-achieving codes presented in \cite{Silas2019}, which employ symbol-wise decode-and-forward in the three-node network. It is known from \cite{Silas2019} that there exists an $(N_1, N_2, N_3)$-achievable code with the following parameters: 
\begin{align}
    n_1 &= T + 1 - N_3, \quad
    n_2 = 0, \quad
    n_3 = T + 1 - N_1 \nonumber \\
    k_1 &= T + 1 - N_1 - N_3, \quad 
    k_2 = 0. \label{eq:singleuser1}
\end{align}
Similarly, there exists an $(N_1, N_2, N_3)$ achievable code with the following parameters:
    \begin{align}
    n_1' &= 0, \quad
    n_2' = T + 1 - N_3, \quad
    n_3' = T + 1 - N_2 \nonumber \\
    k_1' &= 0, \quad
    k_2' = T + 1 - N_2 - N_3. \label{eq:singleuser2}
\end{align}
Then, it follows from Lemma~\ref{lemma:timesharing}, that we can concatenate both single-user codes and obtain another $(N_1, N_2, N_3)$-achievable streaming code. This is the construction we denote as Concatenated Symbol-wise Decode-and-forward. 

\begin{lemma} \label{lemma:cswdf}
    For any $A, B \in \mathbb{Z}_+$ and $T \geq N_1 + N_3$, there exists an $(N_1, N_2, N_3)$-achievable code with rate pair
    \begin{align}
        n &= \max( A n_1 + B n_1', A n_2 + B n_2', A n_3 + B n_3') \\
        R_1 &= \frac{A k_1 + B k_1'}{n} \\
        R_2 &= \frac{A k_2 + B k_2'}{n}
    \end{align}
    where $k_1$, $k_1'$, $k_2$, $k_2'$, $n_1$, $n_1'$, $n_2$, $n_2'$, $n_3$ and $n_3'$ are as in \eqref{eq:singleuser1} and \eqref{eq:singleuser2}.
\end{lemma}
\begin{IEEEproof}
    The result follows directly from applying Lemma~\ref{lemma:timesharing} to the capacity-achieving single-user codes known to be achievable from \cite{Silas2019}, and computing the resulting rate pair of the concatenated code.
\end{IEEEproof}

The following Lemma presents an achievable point which, to the best of our knowledge, is the best achievable rate for $R_1$ such that $R_2 = C(T - N_3, N_2)$, i.e., the corner point for maximum $R_2$. This rate can also be obtained using the Concatenated Symbol-wise Decode-and-forward scheme.
\begin{lemma}\label{lemma:uppercorner}
    For $R_2 = C(T - N_3, N_2)$, the following $R_1$ is achievable
    \begin{align}
        R_1 = \frac{(T + 1 - N_3 - N_1)(N_2 - N_3)}{(T + 1 - N_1)(T + 1 - N_3)}
    \end{align}
\end{lemma}
\begin{remark}
    Note that
    \begin{align*}
        R_1 &= \frac{(T + 1 - N_3 - N_1)(N_2 - N_3)}{(T + 1 - N_1)(T + 1 - N_3)} \\&<  \frac{(T + 1 - N_3 - N_2)(N_2 - N_3)}{(T + 1 - N_2)(T + 1 - N_3)} \\&= C(T - N_2, N_3) - C(T - N_3, N_2).
    \end{align*}
    That is, the sum rate at the point from Lemma~\ref{lemma:uppercorner} is strictly lower than the sumrate capacity.
\end{remark}

\subsubsection{Concatenated Message-wise Decode-and-forward}\label{sec:concMW}

Another scheme that is worth analyzing is the Concatenated Message-wise Decode-and-forward. Similar to the CSWDF scheme, we start by finding a single-user achievable streaming code, and then we apply Lemma~\ref{lemma:timesharing}. For this construction, however, we require the delay of all symbols to be equal in each link, that is, the relay must decode the entire source message before it can re-encode it. This effectively separates the communication in two: one with delay $\tilde{T}$ and another with delay $T - \tilde{T}$, where $\tilde{T}$ is a design parameter. 

It then follows that there exists an $(N_1, N_2, N_3)$-achievable code with the following parameters: 
\begin{align}
    n_1 &=  (\tilde{T}_1 + 1)(T - \tilde{T}_1 + 1 - N_3), 
    n_2 = 0, 
    n_3 = (T - \tilde{T}_1 + 1)(\tilde{T}_1 + 1 - N_1) \nonumber \\
    k_1 &= (T - \tilde{T}_1 + 1 - N_3)(\tilde{T}_1 + 1 - N_1), 
    k_2 = 0 \label{eq:msingleuser1}
\end{align}
and similarly, an $(N_1, N_2, N_3)$-achievable code with the parameters
\begin{align}
    n_1' &= 0,
    n_2' = (\tilde{T}_2 + 1)(T - \tilde{T}_2 + 1 - N_3),
    n_3' = (T - \tilde{T}_2 + 1)(\tilde{T}_2 + 1 - N_2) \nonumber \\
    k_1' &= 0,
    k_2' = (T - \tilde{T}_2 + 1 - N_3)(\tilde{T}_2 + 1 - N_2) \label{eq:msingleuser2}
\end{align}
where $\tilde{T}_1 \in \{N_1, \ldots, T - N_3 \}$ and $\tilde{T}_2 \in \{N_2, \ldots, T - N_3 \}$ are design parameters. In particular, one can optimize these parameters in order to find the maximum single-user rate that can be achieved using message-wise decode-and-forward.

\begin{lemma}
    For any $A, B \in \mathbb{Z}_+$ and $T \geq N_1 + N_3$, there exists an $(N_1, N_2, N_3)$-achievable code with rate pair
    \begin{align}
        n &= \max( A n_1 + B n_1', A n_2 + B n_2', A n_3 + B n_3') \\
        R_1 &= \frac{A k_1 + B k_1'}{n} \\
        R_2 &= \frac{A k_2 + B k_2'}{n}
    \end{align}
    where $k_1$, $k_1'$, $k_2$, $k_2'$, $n_1$, $n_1'$, $n_2$, $n_2'$, $n_3$ and $n_3'$ are as in \eqref{eq:msingleuser1} and \eqref{eq:msingleuser2}, substituting $\tilde{T}_i$ with
    \begin{align}
    \tilde{T}_i &= \arg \max_{T'} \min(\tilde{R}_i, \tilde{R}_{bn}) \\
    \tilde{R}_i &= \frac{T' + 1 - N_i}{T' + 1}\\
    \tilde{R}_{bn} &= \frac{T - T' + 1 - N_3}{T - T' + 1}.
\end{align}
\end{lemma}
\begin{IEEEproof}
    As before, this follows directly from applying Lemma~\ref{lemma:timesharing} and computing the rate pair. The message-wise decode-and-forward scheme and its achievability are presented in more detail in \cite{Silas2019}.
\end{IEEEproof}

\subsection{Optimized-Bottleneck Decode-and-Forward} \label{sec:optimized}

While Theorem~\ref{theorem:R2givenR1} is able to achieve the sumrate capacity, and, in that case, the FB-SWDF scheme presented is optimal, there is a significant loss in rate when we can not achieve the upper bound, which can be seen in Fig.~\ref{fig:introresults}, where CSWDF outperforms FB-SWDF for a small portion of the rate region. This is due to our greedy approach in which the rate of the code used from relay to destination is set to the optimistic $R_{bn} = C(T - N_2, N_3)$. Because we are using a high rate code in this link, it translated into a delay spectrum with more symbol with higher delays, which imposes lower rates in the source-to-relay links.

Unfortunately, analytical expressions of the rates as a function of $R_{bn}$, similar to Theorem~\ref{theorem:R2givenR1}, are hard to derive, due to a non-uniformity of the delay spectrum. For that reason, we instead tackle this problem numerically by developing an optimization framework.

Our approach is the following: we will attempt to achieve the highest possible $R_2$ for some given $R_1$. Thus, we start by fixing some $R_1 \in [0, C(T - N_3, N_1)]$. Also, because we define the rate with $n = \max(n_1, n_2, n_3)$, we also simplify the problem, enforcing $n = n_1 = n_2 = n_3$. This is without loss of generality as we can always zero-pad symbols of the higher-rate codes and achieve the same overall rate. This give us the following optimization problem
\begin{align}
    R_2^* = &\max_{R_2, R_{bn}, n} \quad R_2 \\
    &\textrm{s.t.} \quad \mathbf{k}^{(g)}_2 \leq \mathbf{k}^{(g)}_{bn} - \mathbf{k}^{(g)}_1
\end{align}
where $\mathbf{k}^{(g)}_1$, $\mathbf{k}^{(g)}_2$ and $\mathbf{k}^{(g)}_{bn}$ are computed as functions of $R_1$ (which is fixed), $R_2$, $R_{bn}$ and $n$ as in Lemma~\ref{lemma:achiev}. However, the ordering in $\mathbf{k}^{(g)}_{bn}$ is slightly different from how we defined the equally-delayed-symbols grouping description previously. Now, we consider the following ordering:
\begin{align}
    \mathbf{T}^{(g)}_{1} &= \mathbf{T}^{(g)}_2 = [T - N_3, T - N_3 - 1, \ldots, N_2 + 1, N_2] \\
    \mathbf{T}^{(g)}_{bn} &= [ N_3, N_3 + 1, \ldots, T - N_2 - 1, T - N_2].
\end{align}
That is, while the delays for the symbols in the source-to-relay links are ordered in decreasing order, the delays for the relay-to-destination symbols are ordered in increasing order. Note that, by summing the delays, we get an overall delay of $T$. The constraint in our optimization is simply saying that, if the relay-to-destination can transmit $\kappa$ symbols with delay $\tau$, then these same $\kappa$ symbols must be transmitted with delay $T - \tau$ from sources to relay\footnote{The sources could, in principle, transmit with a smaller delay, however, there is no necessity of doing so, thus we keep the formulation simpler by assuming the source will buffer the symbols in order to achieve exactly an overall delay of $T$. In practice, there is no reason to not transmit with the smallest possible delay.}.

Our heuristic approach to solve this optimization problem is described below. First, we choose some large enough $n$ such that for any choice of $R_1$, $R_2$ and $R_{bn}$ we meet the requirement conditions described in Lemma~\ref{lemma:achiev}. Once $R_1$ and $n$ are fixed, the delay spectrum of the code in the link from first source to relay is also fixed.

Then, we initialize the optimization algorithm with $R_{bn} = C(T - N_2, N_3)$, which again (together with $n$) provides us a delay spectrum, and, with the delay spectrum of the first user and the relay codes, we are able to find an achievable $R_2$ through Corollary~\ref{corollary:maxsym}. Note that the outcome of this initial point is exactly Theorem~\ref{theorem:R2givenR1}, since we have chosen to initialize with the same $R_{bn}$. If $R_1 + R_2 = R_{bn}$, that is, if we achieve the sumrate capacity, we stop the optimization algorithm, since nothing else needs to be done. 

However, if $R_1 + R_2 < R_{bn}$, then we need to update the code used from relay to destination. We update the rate $R_{bn}$ using a simple bisection-like algorithm. We start by defining a lower bound $R_\ell^1 = R_1 + R_2 < R_{bn}$ and an upper bound $R_u^1 = R_{bn}$. Then, at the $i$th iteration, we take the following steps
\begin{enumerate}
    \item Update $R_{bn}^{i} = \frac{R_\ell^{i-1} + R_u^{i-1}}{2}$.
    \item Compute the delay spectrum of the relay-to-destination code with rate $R_{bn}^{i}$ using Lemma~\ref{lemma:achiev}.
    \item Compute the delay constraint imposed on the code from the second source to the relay given the delay spectra of the codes from first source to relay and from relay to destination.
    \item Compute $R_2^{i}$ using Corollary~\ref{corollary:maxsym}.
    \item If $R_1 + R_2^{i} = R_{bn}$, we update the lower bound $R_\ell^{i+1} = R_{bn}$ and keep the upper bound $R_u^{i+1} = R_u^{i}$. Otherwise, we update the upper bound $R_u^{i+1} = R_{bn}$ and keep the lower bound $R_\ell^{i+1} = R_\ell^i$. 
    \item If $R_u^{i+1} - R_\ell^{i+1} > \epsilon$, proceed to the $i+1$th iteration. Otherwise, end the algorithm.
\end{enumerate}

\begin{remark}
    {From construction, every step will update either the lower bound (increasing it) or the upper bound (decreasing it). Each bound indeed represents a bound on our algorithm: the lower bound represents an achievable rate (i.e., indeed a lower bound), while the upper bound represents an upper bound on the achievable rate using our framework. Due to the former, the algorithm can only improve upon the FB-SWDF (which is the starting lower bound). Furthermore, the algorithm is guaranteed to eventually converge, as the gap reduces by a factor of $1/2$ every step. }
\end{remark}

Our experiments have shown that this algorithm converges fairly quickly to a high precision degree (e.g. $\epsilon = 10^{-5}$). Further, this algorithm provides the highest achievable rates to the best of our knowledge. The rate outputs of the algorithm can be seen in Section~\ref{sec:results}.

\subsection{Examples} \label{sec:example}

In this example, we use the channel parameters $N_1 = 3$, $N_2 = 2$, $N_3 = 1$, $T = 6$. In order to present numerical values and a code construction, we are going to focus the example in the scenario where the first user wishes to transmit at its single-user capacity, that is, $R_1 = C(T - N_3, N_1) = 0.5$. Then, we wish to construct the codes for each user and the relay in order to maximize the rate achieved by the second user.

\subsubsection{Upper Bound}

By applying Theorem~\ref{upperbound}, we have that $R_2 \leq 0.3$. This not only gives us a benchmark, but is also useful in the code construction of our FB-SWDF, as we will greedly attempt to achieve that rate.

\subsubsection{FB-SWDF}

As explained previously, we start by fixing the rate of the code from relay to destination at $R_{bn} = C(T - N_2, N_3) = 0.8$. We also choose some value of $n$, in this case, $n = 30$. This choice is explained in detail in the proof of Lemma~\ref{lemma:achievablelowercorner}. With the desired rates $R_1$, $R_{bn}$ and the blocklength $n$, we are able to construct codes, as detailed in the proof of Lemma~\ref{lemma:achiev}, that achieve a certain delay spectrum. In this example, the code from relay to destination is able to transmit six symbols with each delay, from 1 to 4, and the code employed by the first user transmits five symbols with each delay from 3 to 5. These codes are presented in Table~\ref{tab:FBSWDF}.

\begin{table*}[]
\centering
\caption{Capacity-achieving FB-SWDF codes} \label{tab:FBSWDF}
\begin{subtable}{\linewidth} \centering
\caption{Source to Relay codes}
\begin{tabular}{|c|c|c|c|c|c|c|}
\hline
\multicolumn{7}{|c|}{User 1: Source to relay}                                                                                                                                                                                      \\ \hline
\backslashbox{Symbol}{Time} & t          & t+1            & t+2                       & t+3                                       & t+4                                           & t+5                                             \\ \hline
  $x^{(1)}_{t, 1}[1] = s_{t, 1}[1]$                         & $a_t[1:5]$ & $a_{t+1}[1:5]$ & $a_{t+2}[1:5]$            & $a_{t+3}[1:5]$                            & $a_{t+4}[1:5]$                                & $a_{t+5}[1:5]$                                  \\ \hline
    $x^{(1)}_{t, 1}[2] = s_{t, 1}[2]$                       & $b_t[1:5]$ & $b_{t+1}[1:5]$ & $b_{t+2}[1:5]$            & $b_{t+3}[1:5]$                            & $b_{t+4}[1:5]$                                & $b_{t+5}[1:5]$                                  \\ \hline
     $x^{(1)}_{t, 1}[3] = s_{t, 1}[3]$                      & $c_t[1:5]$ & $c_{t+1}[1:5]$ & $c_{t+2}[1:5]$            & $c_{t+3}[1:5]$                            & $c_{t+4}[1:5]$                                & $c_{t+5}[1:5]$                                  \\ \hline
    $x^{(1)}_{t, 1}[4]$                       &            & $c_t[1:5]$     & \makecell[c]{$b_t[1:5]$ \\$+ c_{t+1}[1:5]$} & \makecell[c]{ $a_t[1:5]$ \\$+ b_{t+1}[1:5]$ \\ $+ c_{t+2}[1:5] $} & \makecell[c]{ $a_{t+1}[1:5]$ \\$+ b_{t+2}[1:5]$\\ $+ c_{t+3}[1:5] $} & \makecell[c]{ $a_{t+2}[1:5]$ \\$+ b_{t+3}[1:5]$\\ $+ c_{t+4}[1:5] $}   \\ \hline
    $x^{(1)}_{t, 1}[5]$                       &            &                & $c_t[1:5]$                & \makecell[c]{$b_t[1:5]$ \\$+ 2c_{t+1}[1:5]$}                & \makecell[c]{$a_t[1:5]$\\ $+ 2b_{t+1}[1:5]$ \\$+ 3c_{t+2}[1:5] $}   & \makecell[c]{$a_{t+1}[1:5]$\\ $+ 2b_{t+2}[1:5]$ \\$+ 3c_{t+3}[1:5] $} \\ \hline
       $x^{(1)}_{t, 1}[6]$                    &            &                &                           & $c_t[1:5]$                                & \makecell[c]{$b_t[1:5]$ \\$+ 3c_{t+1}[1:5]$}                    & \makecell[c]{$a_t[1:5]$ \\$+ 3b_{t+1}[1:5]$ \\$+ 4c_{t+2}[1:5] $}     \\ \hline
\multicolumn{7}{|c|}{User 2: Source to relay}                                                                                                                                                                                      \\ \hline
         $x^{(1)}_{t, 2}[1] = s_{t, 2}[1]$                  & $d_t[1:9]$ & $d_{t+1}[1:9]$ & $d_{t+2}[1:9]$            & $d_{t+3}[1:9]$                            & $d_{t+4}[1:9]$                                & $d_{t+5}[1:9]$                                  \\ \hline
          $x^{(1)}_{t, 2}[2]$                 &            & $d_t[1:9]$     & $d_{t+1}[1:9]$            & $d_{t+2}[1:9]$                            & $d_{t+3}[1:9]$                                & $d_{t+4}[1:9]$                                  \\ \hline
            $x^{(1)}_{t, 2}[3]$               &            &                & $d_t[1:9]$                & $d_{t+1}[1:9]$                            & $d_{t+2}[1:9]$                                & $d_{t+3}[1:9]$                                  \\ \hline
\end{tabular}
\end{subtable}

\vspace{.2cm}

\begin{subtable}{\linewidth}\centering
\caption{Relay to Destination codes}
\begin{tabular}{|c|c|c|c|c|c|}
\hline
\multicolumn{6}{|c|}{Relay to Destination}                                                                                                                                                                                                                                                                                                                                                          \\ \hline
\backslashbox{Symbol}{Time} & $t+2$      & $t+3$                       & $t+4$                         & $t+5$                            & $t+6$                                                                                                                                                                                                                                     \\ \hline
$x^{(2)}_t[1]$                                 & $d_t[1:6]$ & $d_{t+1}[1:6]$              & $d_{t+2}[1:6]$                & $d_{t+3}[1:6]$                   & $d_{t+4}[1:6]$                                                                                                                                                                                                                            \\ \hline
$x^{(2)}_t[2]$                                 &            & $c_{t}[1:5]\frown d_{t}[7]$ & $c_{t+1}[1:5]\frown d_{t}[7]$ & $c_{t+2}[1:5]\frown d_{t}[7]$    & $c_{t+3}[1:5]\frown d_{t}[7]$                                                                                                                                                                                                             \\ \hline
$x^{(2)}_t[3]$                                 &            &                             & $b_{t}[1:5] \frown d_{t}[8]$  & $b_{t+1}[1:5] \frown d_{t+1}[8]$ & $b_{t+2}[1:5] \frown d_{t+2}[8]$                                                                                                                                                                                                          \\ \hline
$x^{(2)}_t[4]$                                 &            &                             &                               & $a_{t}[1:5] \frown d_{t}[9]$     & $a_{t+1}[1:5] \frown d_{t+1}[9]$                                                                                                                                                                                                          \\ \hline
$x^{(2)}_t[5]$                                 &            &                             &                               &                                  & \makecell[c]{$d_t[1:6]$ \\ $+ c_t[1:5] \frown d_t[7]$ \\ $+ b_t[1:5] \frown d_t[8]$ \\ $+a_t[1:5] \frown d_t[9]$} \\ \hline
\end{tabular}
\end{subtable}
\end{table*}

Now, one can see that the symbols transmitted with delay 5 (from the first user to the relay) must be matched with the symbols transmitted with delay 1 (from relay to destination) in order to achieve an overall delay of $T = 6$. Similarly, symbols transmitted with delay 4 will be matched with symbols transmitted with delay 2, and so on. Therefore, the second user has the following ``budget'': it can transmit six symbols with delay 2 (to be matched with the remaining six symbols with delay 4), one symbol with delay 3 (to be matched with the remaining one symbol with delay 3), one symbol with delay 4, and one symbol with delay 5. This gives us the following tuple
\begin{align}
    \mathbf{T}^{(g, 1)}_2 = [5, 4, 3, 2] \\
    \mathbf{k}^{(con)} = [1, 1, 1, 6].
\end{align}
By applying Corollary~\ref{corollary:maxsym}, we find out that we can, in fact, obtain a code with $k^{(1)}_2 = 9$. In fact, it turns out that we can transmit all 9 symbols with delay 2 by simply repeating a diagonally-interleaved MDS code with rate $2/3$. {However, since our goal is to recover with delay $T = 4$, we can choose to wait (i.e., buffer) some symbols, so all symbols are recovered by the destination at the same time instant.} This code is also presented in Table~\ref{tab:FBSWDF}. The matching of delay spectra can be seen in Table~\ref{tab:delays}. Recall that $R_2 = 9/30 = 0.3$ is exactly the upper bound. 
\begin{table}[]
\centering
\caption{Number of symbols transmitted with each delay in each hop. \textcolor{blue}{Blue} symbols are transmitted by the first user, while \textcolor{green}{green} symbols are transmitted by the second user.} \label{tab:delays}
\begin{tabular}{|c|c|c|c|c|}
\hline
\backslashbox{$\mathbf{T}$ in the first hop}{$\mathbf{T}$ in the second hop} & 1  & 2  & 3  & 4  \\ \hline
2                                                                                 &  &  &  & \textcolor{green}{6} \\ \hline
3                                                                                 &    &    & (\textcolor{blue}{5}, \textcolor{green}{1}) &    \\ \hline
4                                                                                 &    & (\textcolor{blue}{5}, \textcolor{green}{1}) &    &    \\ \hline
5                                                                                 & (\textcolor{blue}{5}, \textcolor{green}{1}) &    &    &    \\ \hline
\end{tabular}
\end{table}

\subsubsection{CSWDF}

For CSWDF, one way to solve this problem is by optimizing $A$ and $B$ such that it results in $R_1 = 0.5$ and maximizes $R_2$. That is, under CSWDF, we wish to solve the following optimization problem
\begin{align}
    R_2^* = \max_{A, B} \quad &R_2 \\
     \textrm{s.t.~} \quad &R_1 = 0.5 \\
     &n = \max(6A, 6B, 4A + 5B) \\
     &R_1 = \frac{3A}{n} \\
     &R_2 = \frac{4B}{n}
\end{align}
where the constraints come from Lemma~\ref{lemma:cswdf}. Solving this optimization leads to $A = 5$ and $B = 2$, which results in $n = 30$, $k_1 = 15$ and $k_2 = 8$, i.e., $R_1 = 0.5$ as desired, and $R_2 = 0.2667$, slightly lower than the upper bound and the rate achieved by our FB-SWDF scheme.

\subsubsection{MSWDF}

Similar to the CSWDF, this requires solving an optimization problem, however, for MSWDF we also need to optimize the parameters $\tilde{T}_i$, $i \in \{1, 2\}$. Doing so results in $B = 0$ and any $A$, that is, in order to achieve $R_1 = 0.5$, we need $R_2 = 0$ for MSWDF. 

\subsubsection{Comparison}

We hope that this example demonstrates the considerably suboptimality of MSWDF and CSWDF. The major difference between our scheme and CSWDF, fundamentally, is that our scheme allows for ``merging'' of the streams arriving at the relay from different sources, encoding all streams within the same code. One main contribution of this paper is to present a systematic way of designing such ``merging'' in a delay-spectrum efficient manner, through the framework provided by Lemma~\ref{lemma:achiev} and Corollary~\ref{corollary:maxsym}.

\section{Results}\label{sec:results}

\begin{figure}[h]
	\centering
	\includegraphics[draft=false,width=\linewidth]{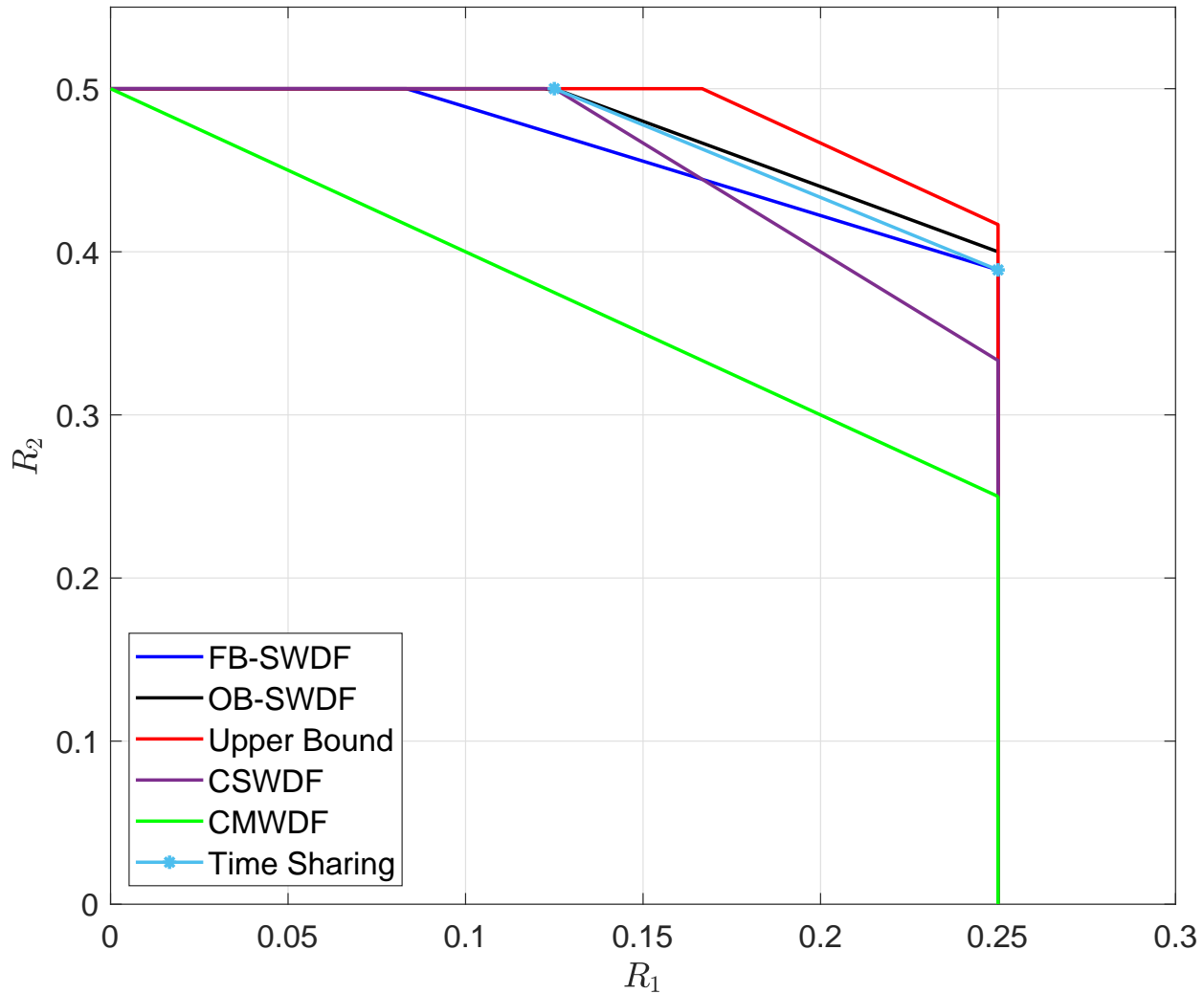}
	\caption{Rate region for the weak relay-destination bottleneck regime with parameters $N_1=3,N_2=2,N_3=1,T=4$.}
	\label{fig:fig1}
\end{figure}

In this section, we present the rate region for two different scenarios. In both cases, the curves are labeled according to the scheme presented in each respective section previously. The curve labeled as ``Time Sharing'' in Fig~\ref{fig:fig1} represents the best achievable region for which we have closed form expressions.

\begin{figure}[h]
	\centering
	\includegraphics[draft=false,width=\linewidth]{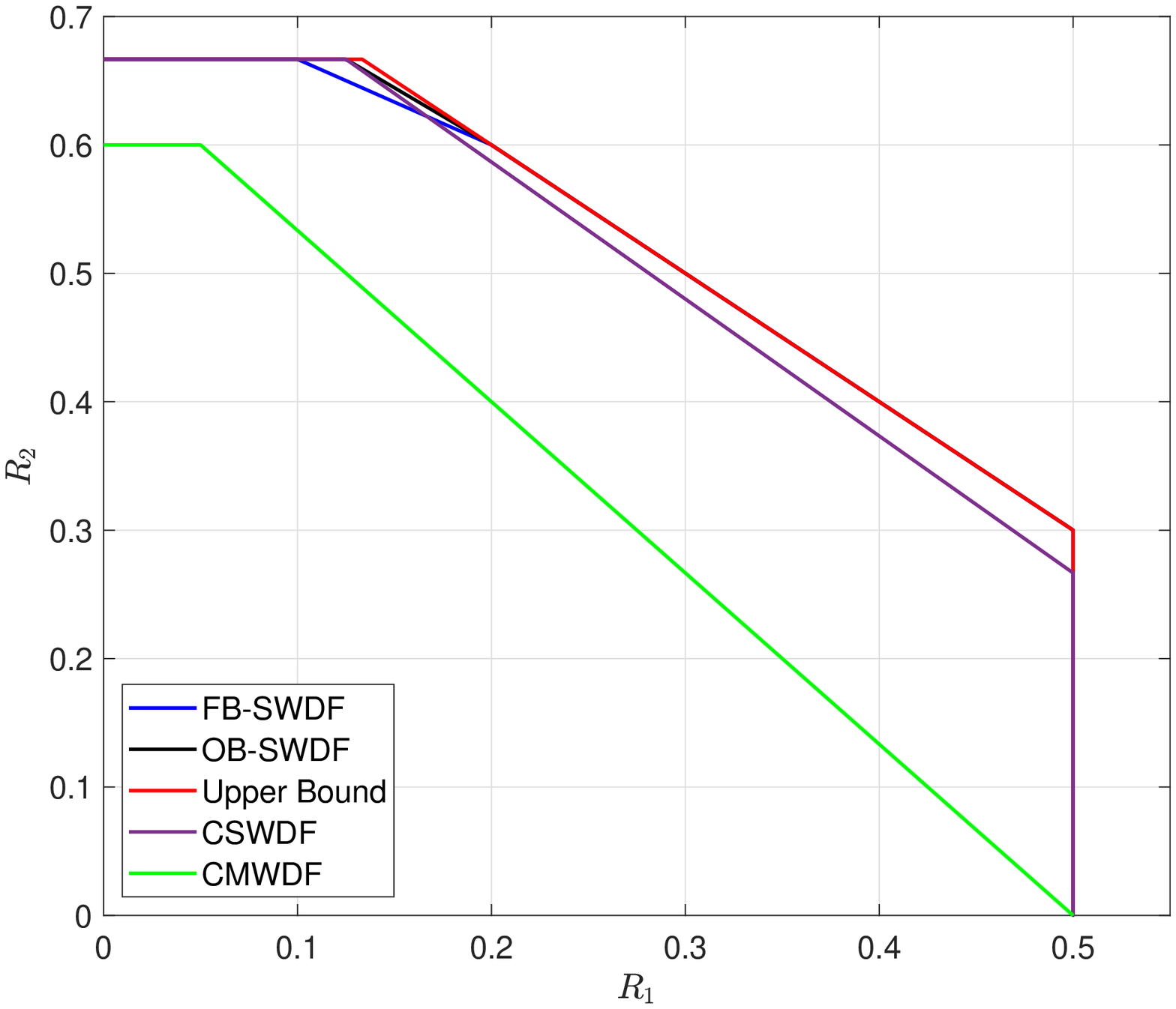}
	\caption{Rate region for the weak relay-destination bottleneck regime with parameters $N_1=3,N_2=2,N_3=1,T=6$.}
	\label{fig:fig2}
\end{figure}

In the first scenario, we have a small $T$, such that the condition in Corollary~\ref{corollary:minimumT} is not met. This is presented in Fig.~\ref{fig:fig1}. Note that, in this case, no scheme is able to achieve the sumrate. 
By slightly increasing $T$, we are able to achieve the sumrate capacity in a noticeable part of the capacity region, which is shown in Fig.~\ref{fig:fig2}. Note that, in this case, the message-wise scheme is unable to achieve even the single-user capacity for $R_2$, which is also shown in \cite{Silas2019}.

Furthermore, in both cases, it can be seen that CSWDF is able to achieve the optimal point for maximal $R_2$. This has been observed in all settings we have experimented, although it remains to be proven. Nonetheless, it suggests that the ``Time Sharing'' scheme represents a good achievable rate, for which we have closed form expressions. In fact, in Fig.~\ref{fig:fig2}, it achieves the same performance as OB-SWDF.

Although our schemes are unable to always achieve the sumrate, and unable to achieve the whole rate region, comparing them to the alternative---CMWDF---should show that the proposed schemes are significantly superior.

\section{Extension to General Parameters}\label{sec:allregimes}

In this section, we extend the previous achievability results for networks with general parameters.
In order to do such analysis, we consider four classifications of network based on the upper bound.

First, we separate the regimes of operation based on the sumrate bottleneck. If $C(T - N_2, N_3) > C(T - N_3, N_1) + C(T - N_3, N_2)$, we denote the regime as a ``source-relay bottleneck'', otherwise, we denote it as ``relay-destination bottleneck''. Further, we also consider how ``strong'' the bottleneck is.

For the source-relay bottleneck regimes, we denote it as a ``strong'' bottleneck if we also have $C(T - N_1, N_3) \geq C(T - N_3, N_1) + C(T - N_3, N_2)$. (Recall that $C(T - N_2, N_3) \geq C(T - N_1, N_3)$ since $N_1 \geq N_2$.) This regime represents a scenario where the channel condition in the link from relay to destination is fairly better than both links from sources to relay. We show that, in this ``strong'' regime, both users can transmit at their single-user capacity.

In the ``weak'' source-relay bottleneck, on the other hand, we could not obtain an achievable scheme in which both users transmit at their single-user capacity, even though that is the upper bound. However, we derive a condition for the achievability.

As for the ``relay-destination'' bottleneck regime, the link from relay to destination acts as a bottleneck, and both users can not transmit at their single-user capacity. In the paper, we have analyzed the scenario where $N_2 \geq N_3$, which we denote as ``weak'' relay-destination bottleneck. In this section, we further analyze the strong relay-destination bottleneck, in which the sumrate can be achieved by single-user communication (i.e., it can be achieved with $R_1 = 0$). We further show that the entire capacity region is achievable in this regime.

Below, we itemize the four regimes and the upper bound conditions that characterize each one.

\begin{itemize}
    \item Strong source-relay bottleneck: $C(T - N_1, N_3) \geq C(T - N_3, N_1) + C(T - N_3, N_2)$.
    \item Weak source-relay bottleneck: $C(T - N_2, N_3) \geq C(T - N_3, N_1) + C(T - N_3, N_2) > C(T - N_1, N_3)$.
    \item Weak relay-destination bottleneck: $C(T - N_3, N_1) + C(T - N_3, N_2) \geq C(T - N_2, N_3) \geq C(T - N_3, N_2)$.
    \item Strong relay-destination bottleneck: $C(T - N_3, N_1) \leq C(T - N_2, N_3) \leq C(T - N_3, N_2)$.
\end{itemize}

\subsection{Strong source-relay bottleneck}

In this case, the converse region is entirely defined by
\begin{align}
    R_1 &\leq C(T - N_3, N_1) \\
    R_2 &\leq C(T - N_3, N_2)
\end{align}
since the sumrate bound is never active. Furthermore, it is easy to see that $N_2 \geq N_3$ is required in such scenario, otherwise, it is impossible for the condition for strong source-relay bottleneck to hold (that is, otherwise we would be in the strong relay-destination bottleneck).

In order to show that the entire rate region is achievable, we state the following Lemma
\begin{lemma}
    For the strong source-relay bottleneck regime, there exists an $(N_1, N_2, N_3)$-achievable code with the following rate pair
    \begin{align}
        R_1 &= C(T - N_3, N_1) \\
        R_2 &= C(T - N_3, N_2)
    \end{align}
\end{lemma}
\begin{IEEEproof}
    This follows from Lemma~\ref{lemma:cswdf} (Concatenated Symbol-wise Decode-and-forward scheme), using $A = B = 1$. This will give us an $(N_1, N_2, N_3)$-achievable code with the following parameters: $k_1 = (T + 1 - N_1 - N_3)$, $k_2 = (T + 1 - N_2 - N_3)$, $n_1 = n_2 = (T + 1 - N_3)$, $n_3 = (T + 1 - N_1) + (T + 1 - N_2)$.  Then, it remains to show that $n = n_1 = n_2 \geq n_3$. 
    
    The proof is as follows: since $N_1 \geq N_2$, we have
    \begin{align}
        \frac{T + 1 - N_1 - N_3}{T + 1 - N_1} \leq \frac{T + 1 - N_2 - N_3}{T + 1 - N_2}. \label{eq:concatenation_inequality1}
    \end{align}
    From \eqref{eq:concatenation_inequality1}, it follows that
    \begin{align}
        \frac{(T + 1 - N_1 - N_3) + (T + 1 - N_2 - N_3)}{(T + 1 - N_1) + (T + 1 - N_2)} \geq \frac{T + 1 - N_1 - N_3}{T + 1 - N_1}. \label{eq:concatenation_inequality2}
    \end{align}
    Further, from the strong source-relay bottleneck assumption, we have
    \begin{align}
        \frac{T + 1 - N_1 - N_3}{T + 1 - N_1} \geq \frac{(T + 1 - N_3 - N_1) + (T + 1 - N_3 - N_2)}{T + 1 - N_3}.
    \end{align}
    It follows directly then that
    \begin{align}
        (T + 1 - N_1) + (T + 1 - N_2) \leq (T + 1 - N_3).
    \end{align}
    Now recall that, for $A = B = 1$, we have $n_1 = n_2 = T + 1 - N_3$ and $n_3 = (T + 1 - N_1) + (T + 1 - N_2)$, therefore, we completed the proof that $n = n_1 = n_2 \geq n_3$.
\end{IEEEproof}

Furthermore, since each single-user capacity point is achievable, by applying Lemma~\ref{lemma:timesharing}, it is easy to see that the entire capacity rate region is achievable using CSWDF.

\begin{figure}
    \centering
    \includegraphics[draft=false,width=\linewidth]{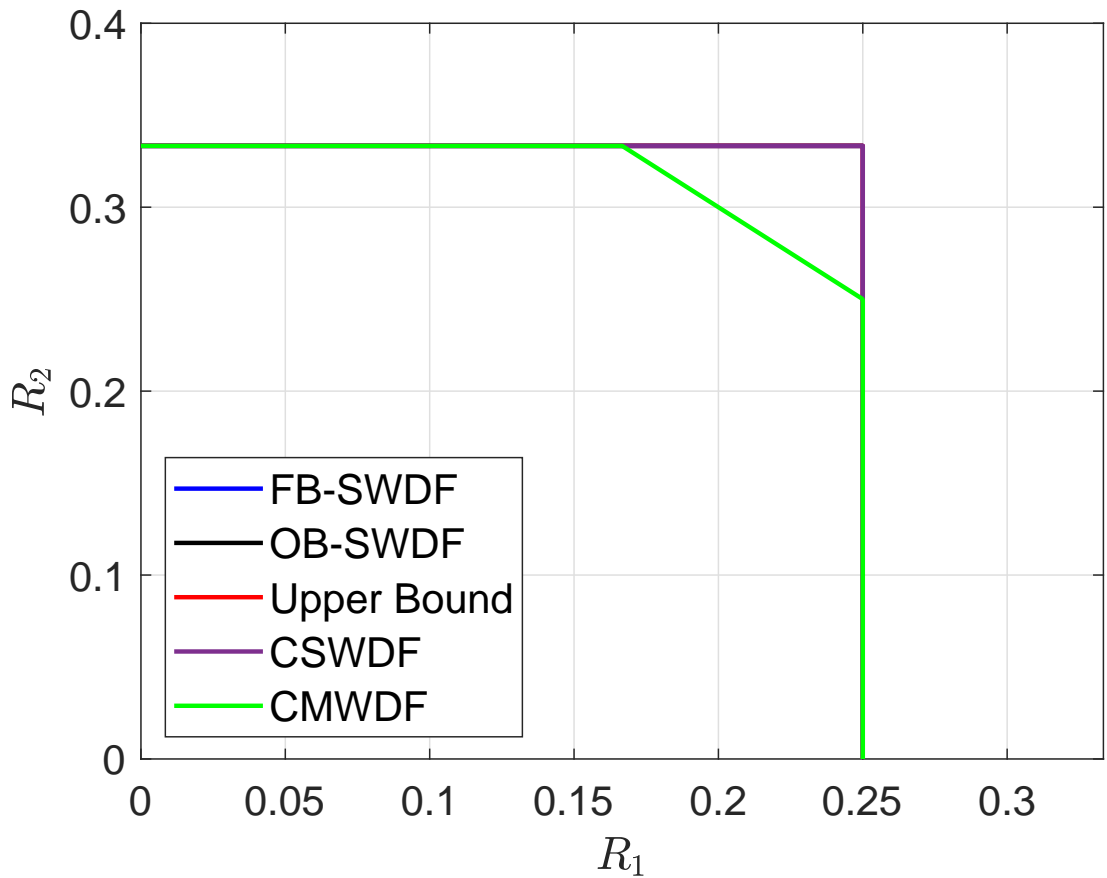}
    \caption{Capacity Rate Region for a strong source-relay bottleneck regime with parameters $N_1 = 9$, $N_2 = 8$, $N_3 = 1$, $T = 12$. All schemes other than CMWDF are able to achieve the upper bound.}
    \label{fig:strongSR}
\end{figure}

An example of such rate region is presented in Fig.~\ref{fig:strongSR}, using the parameters $N_1 = 9$, $N_2 = 8$, $N_3 = 1$, $T = 12$. Note that, while the proof we used for achievability uses the concatenated symbol-wise scheme for simplicity, our proposed schemes are also able to achieve the entire rate region, while MSWDF is not guaranteed to.

\subsection{Weak source-relay bottleneck}

\begin{figure}
    \centering
    \includegraphics[draft=false,width=\linewidth]{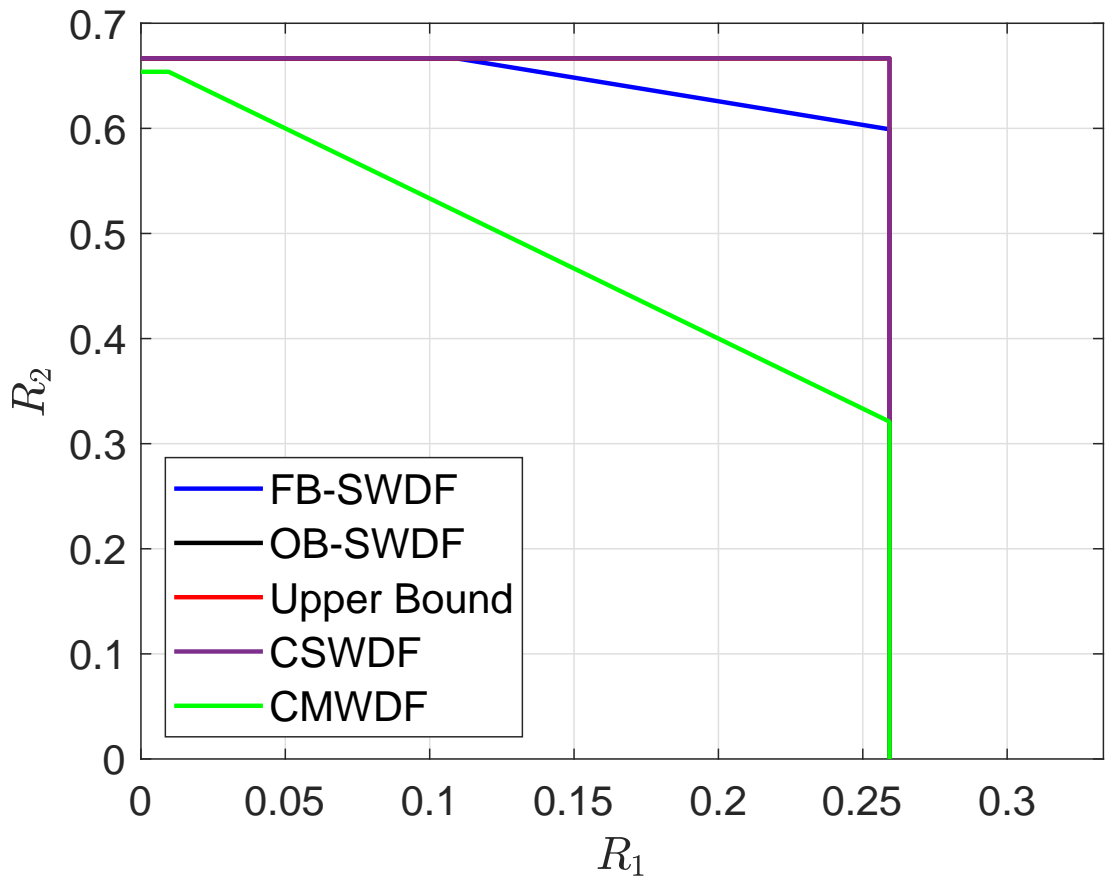}
    \caption{Capacity Rate Region for a weak source-relay bottleneck regime with parameters $N_1 = 20$, $N_2 = 9$, $N_3 = 1$, $T = 27$. }
    \label{fig:weaksr1}
\end{figure}

\begin{figure}
    \centering
    \includegraphics[draft=false,width=\linewidth]{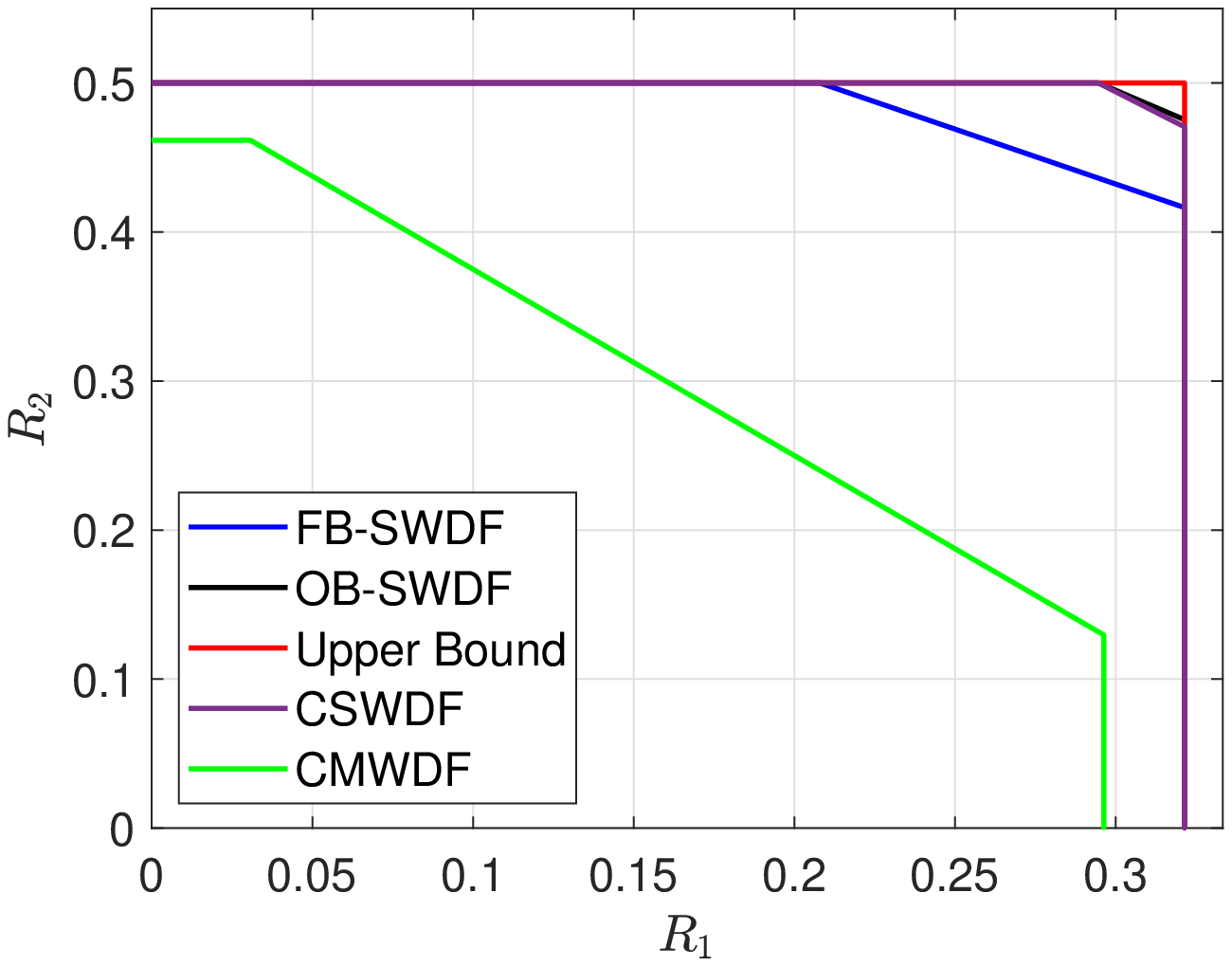}
    \caption{Capacity Rate Region for a weak source-relay bottleneck regime with parameters $N_1 = 19$, $N_2 = 14$, $N_3 = 3$, $T = 30$.}
    \label{fig:weaksr2}
\end{figure}

In this case, similar to before, the converse is reduced to
\begin{align}
    R_1 &\leq C(T - N_3, N_1) \\
    R_2 &\leq C(T - N_3, N_2).
\end{align}
However, we may not be able to achieve the corner point (i.e., achieve this bound with equality). Nonetheless, all schemes presented in our paper can be employed in this regime, although the FB-SWDF scheme is certainly going to outperform if we choose to keep the chosen $R_{bn}$, as it is certainly not achievable in this regime. We take this opportunity to highlight the effectiveness of OB-SWDF, which again presents the best achievability in Fig.~\ref{fig:weaksr1} and Fig.~\ref{fig:weaksr2}.

Nonetheless, we can guarantee achievability of the entire region under the following condition.
\begin{lemma} \label{lemma:weaksr}
    For a network in the weak source-relay bottleneck regime, if
    \begin{equation*}
        T \leq N_1 + N_2 - N_3 - 1
    \end{equation*}
    then the entire rate region is achievable.
\end{lemma}
\begin{IEEEproof}
    This follows directly from Lemma~\ref{lemma:cswdf} with $A = B = 1$. If the condition stated in the lemma holds, we are able to achieve the corner point $R_1 = C(T - N_3, N_1)$ and $R_2 = C(T - N_3, N_2)$, and, again, since the single-user points are achievable, we can achieve the entire rate region.
\end{IEEEproof}
\begin{remark}
    The Lemma can be seen in the example figures. In the first case, we have $T = 27 \leq 27 = N_1 + N_2 - N_3 - 1$, and we can achieve the capacity region. In the second example, we have $T = 30 > 29 = N_1 + N_2 - N_3 - 1$, and we can not achieve the capacity region.
\end{remark}

\subsection{Strong relay-destination bottleneck}

The condition for the network to be in this regime of operation is that $N_3 \geq N_2$. In this case, we have that the converse is given by
\begin{align}
    R_1 &\leq C(T - N_3, N_1) \\
    R_1 + R_2 &\leq C(T - N_2, N_3).
\end{align}
We now show that both points are achieved with equality.
\begin{lemma}
    In the strong relay-destination bottleneck regime, the following rate pairs are achievable
    \begin{align}
        R_1 = 0, R_2 = C(T - N_2, N_3) \\
        R_1 = C(T - N_3, N_1), R_2 = C(T - N_2, N_3) - C(T - N_3, N_1).
    \end{align}
\end{lemma}
\begin{IEEEproof}
    The rate pair $R_1 = 0, R_2 = C(T - N_2, N_3)$ is achievable using the three-node network capacity-achieving codes presented in \cite{Silas2019}, or, in other words, a single-user capacity achieving code. The second rate pair can be easily achieved using our FB-SWDF scheme. In order to see that, note that these expressions are the same as Lemma~\ref{lemma:achievablelowercorner}, and therefore, the rate pair is achievable under the condition described in Corollary~\ref{corollary:minimumT}. That is, we need
    \begin{equation}
        (T + 1 - N_3)(T + 1 - N_1 - N_2) \geq (T + 1 - N_1 - N_3)(N_2 - N_3).
    \end{equation}
    However, note that from the strong relay-destination bottleneck assumption, we have $N_3 \geq N_2$, therefore, $N_2 - N_3 \leq 0$, thus the right-hand side is negative. The left-hand side is always positive, as $T \geq N_1 + N_3$ (otherwise we have $R_1 = 0$ and this degenerates to a single-user setting). Therefore, the condition of the corollary is trivially met, completing the proof.
\end{IEEEproof}

It then easily follows that the entire capacity region is achievable from Lemma~\ref{lemma:timesharing}.

\begin{figure}
    \centering
    \includegraphics[draft=false,width=\linewidth]{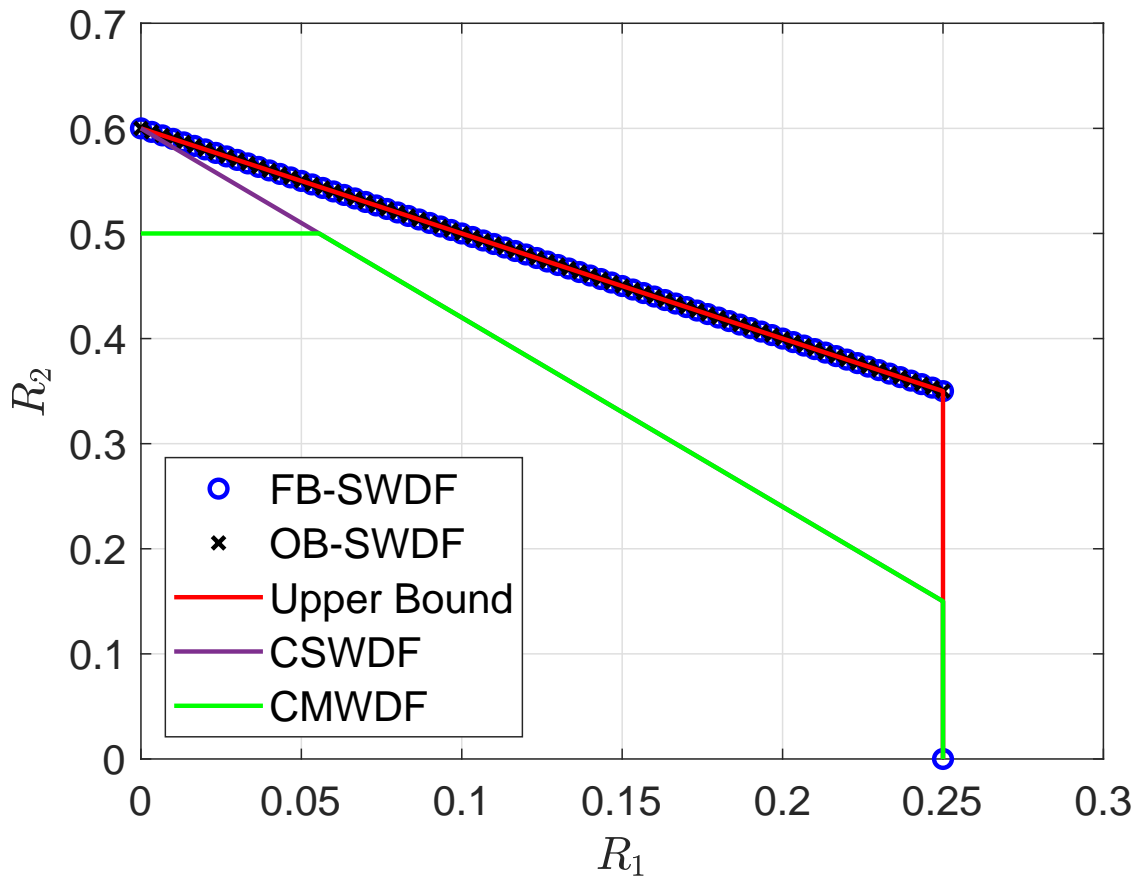}
    \caption{Capacity Rate Region for a strong relay-destination bottleneck regime with parameters $N_1 = 3$, $N_2 = 1$, $N_3 = 2$, $T = 5$.}
    \label{fig:strongRD}
\end{figure}

An example of such rate region is presented in Fig.~\ref{fig:strongRD}. Note that neither the concatenated symbol-wise decode-and-forward or the concatenated message-wise decode-and-forward are able to achieve the entire capacity region, while our proposed scheme is able to.

\section{Comments and Limitations}

In this section we wish to make some comments about our results, acknowledge a few limitations and discuss future work.

One comment to be made about our results is about the asymptotic optimality. For a sufficiently large $T$, it is easy to see that the condition on Corollary~\ref{corollary:minimumT} is always met, as the left hand side grows with $T^2$, while the right hand side grows with $T$. Furthermore, the upper corner point tends to degenerate to the single-user scenario with $T \to \infty$, as can be seen from the upper bound and from Lemma~\ref{lemma:uppercorner}. Therefore, with $T \to \infty$, the capacity region is entirely achievable. 

In terms of the presented schemes, our OB-SWDF presents two limitations: one is that the results are numerical, thus no performance guarantees are given. However, one may construct the optimization in such a way that it certainly passes through the rates achieved by FB-SWDF and CSWDF, guaranteeing an outperformance of these schemes. The second limitation is that the choice of $n$ in order to obtain the best possible performance might be very large (e.g. $n \geq 2000$) and impractical for the applications which are the target of low latency streaming codes. This is highly dependent on the channel parameters. However, experimental results show that one may achieve rates close to the presented in our results even with a significantly smaller choice of $n$.

Another limitation is that we focus on time-invariant codes, and, more specifically, our relaying strategy does not exploit the knowledge of the erasures from source to relay. This constraint on the streaming codes reduces the capacity. However, this is justified by practical limitations on the relay's processing capabilities, as well as by allowing no-overhead transmission. Nonetheless, it is possible that the use of adaptive schemes may allow for better performance in the multi-access setting and it is considered as future work.

{Finally, it should be noted that while we focus our analysis on the uplink, our framework is fairly symmetrical and can be simply generalized to the downlink, in which a source wishes to transmit different messages to different destinations (assuming wired connections between nodes). This is especially useful since many applications, such as cloud gaming, involve two-way communication.}

\section{Conclusion}

In this paper, we have presented an in-depth analysis of achievability using symbol-wise decode-and-forward. We have derived the condition for which the sumrate is achievable, and, when it is, what part of the whole rate region is achievable. Furthermore, we have compared the proposed schemes to concatenated message-wise decode-and-forward, which is the naive solution to the problem, and the performance of our scheme is clearly superior. It should also be noted that, while the analysis in this paper is restricted to 2 users, the tools and schemes presented can be easily extended to any number of users, and the achievable rates can easily be solved numerically, although analysis of the closed-form expressions become increasingly hard to derive.

For future works, the gap between achievability and converse should be tightened. It is likely that the current converse is optimistic, and new converse-deriving techniques are necessary in order to obtain better results. It is also unclear whether employing a different (than symbol-wise decode-and-forward) strategy in the relay can improve the achievable results, and an analysis based on information rates could be helpful.

\newpage

\appendix
\newcommand{\bs}{\mathbf{s}}
\newcommand{\bT}{\mathbf{T}}
\newcommand{\bN}{\mathbf{N}}
\begin{IEEEproof}[Delay Spectrum of Symbol-wise Decode-and-Forward]
     Let
	\begin{equation}
	\bs''_t = [\bs_{t - (T^{(1)}[1] )}[1], \bs_{t - (T^{(1)}[2])}[2], \ldots, \bs_{t - (T^{(1)}[k])}[k]]
	\end{equation}
	that is, $\bs''_t$ is the vector of symbols that are decoded at time $t$. From assumption, the relay has access to this vector at time $t$. The relay can then apply the permutation
	\begin{equation}
	\bs'_t = \pi \bs''_t
	\end{equation}
	and encode such sequence. Now, consider the $j$th symbol. It has been recovered with a delay $T^{(1)}[j]$ by the relay, then it has been permuted into the $j'$ symbol and recovered with delay $T^{(2)}[j']$ by the destination. For all $j$ and all permutations, the destination has recovered the symbol by time $T^{(1)}[j] + T^{(2)}[j']$, where $j'$ is defined by the permutation.
	
	That is, the delay spectrum of symbol-wise decode-and-forward is given by $\mathbf{T}^{(1)} + \pi \mathbf{T}^{(2)}$ where $\pi$ is some permutation.
\end{IEEEproof}

\begin{IEEEproof}[Proof of Lemma~\ref{lemma:conc}]
	Consider the concatenation code of both codes. Since the first code is able to achieve delay spectrum $\bT'$ under $\bN$ erasures, the first $k'$ symbols can be recovered certainly at times $\bT'$. Similarly, since the second code achieves delay spectrum $\bT''$ under $\bN$ erasures, the last $k''$ symbols can be recovered certainly at times $\bT''$. Thus, the concatenation code achieves the delay spectrum $[\bT', \bT'']$ under $\bN$ erasures.
\end{IEEEproof}

\begin{IEEEproof}[Proof of Lemma~\ref{lemma:permutation}]
    This follows directly from the choice of encoder. Let us denote by $f_{t}(\{\bs_j\}_{j=0}^{t}$ the encoding function of a code that achieves delay spectrum $\bT$ under $N$ erasures. Without loss of generality, let us analyze the first two source symbols. This code is able to recover them by times $T[1]$ and $T[2]$, respectively. Now, consider the code generated by applying the function to $\bs'$, i.e., $f_{t}(\{ \bs'_j\}_{j=0}^{t})$, where $\bs' = \pi \bs$. Again without loss of generality, assume the permutation swaps the first two positions. Then, $\bs'[1]$ is now recovered at time $T[1]$, and $\bs'[2]$ is recovered at time $T[2]$. However, $\bs'[1] = \bs[2]$, therefore, $\bs[2]$ is recovered at time $T[1]$, and similarly for $\bs[1]$.
    
    It is easy to see that, in general, applying the desired permutation over the source symbols before encoding results in the desired permuted delay spectrum. Thus, if a code is able to achieve delay spectrum $\bT$ under $N$ erasures, permuting the source symbols and using the same code suffices to acheive delay spectrum $\pi \bT$ under $N$ erasures.
\end{IEEEproof}

The following lemma has been proven in \cite{Silas2019} and is useful in the proof of Lemma~\ref{lemma:achiev}.
\begin{lemma}\label{lemma:intMDS}
	A systematic $(N+k, k, \bT)_{\mathbb{F}}$ point-to-point diagonally-interleaved MDS code \cite{badr2013streaming} achieves the delay spectrum $\bT = [N, \ldots, N + k - 1]$ under $N$ erasures.
\end{lemma}

\begin{IEEEproof}[Proof of Lemma~\ref{lemma:achiev}]
    Recall that we wish to construct an $(n, k)$ code such that, under $N$ erasures, $k^{(g)}[j]$ symbols are recovered at time $T^{(g)}[j]$ and
    \begin{equation}
        k^{(g)}[j] = \begin{cases}
        n - \frac{T^{(g)}[1]}{N}(n - k), &j = 1\\
        \frac{n - k}{N}, &j \geq 2
        \end{cases}.
    \end{equation}
    For the remaining of the proof, we denote $k^{(g)}[j] = \frac{n - k}{N}$, i.e., we assume $j \geq 2$. 

	Consider the following coding scheme: denote by $m = (T^{(g)}[1] + 1 - N)$ the number of possible delays. We concatenate $k^{(g)}[1] = n - \frac{T^{(g)}[1]}{N}(n - k)$ diagonally-interleaved MDS codes with parameters $(N + m, m)$ and  $k^{(g)}[j] - k^{(g)}[1] =  \frac{T^{(g)}[1] + 1}{N}(n - k) - n$ interleaving MDS codes with parameters $(N + m - 1, m - 1)$.
	
	It follows directly from Lemma~\ref{lemma:intMDS} and Lemma~\ref{lemma:conc} that $k^{(g)}[1]$ symbols are recovered with delay $T^{(g)}[1]$ and that $k^{(g)}[j]$ symbols are recovered with every other delay (i.e., $T^{(g)}[j]$). From definition, we have $k = k^{(g)}[1] + \sum_{j=2}^{m} k^{(g)}[j]$, since the number of possible delays is exactly $m$.
	
	Therefore, it remains to show that the number of channel uses this proposed code uses, which we will briefly denote as $n'$, is the designed number of channel uses $n$.
	
	The number of channel uses of this concatenation is, from definition of concatenation, given by
	\begin{align}
	n'= &k^{(g)}[1] (N + m) + (N + m - 1)(k^{(g)}[j] - k^{(g)}[1]) \\
	&= (N + m - 1)k^{(g)}[j] + k^{(g)}[1] \overset{(b)}{=} N k^{(g)}[j] + k \\
	&\overset{(c)}{=} (n - k) + k = n
	\end{align}
	where $(a)$ follows from $k^{(g)}[1] + (m - 1)k^{(g)}[j] = k$ and $(b)$ follows from $k^{(g)}[j] = \frac{1}{N}(n - k)$.
	
	Now, note that this requires $n - \frac{T[1]}{N}(n - k)$ and $\frac{T[1] + 1}{N} (n - k)$ to be integer. Both are true from assumption. This is intuitive from the construction: we are concatenating codes of the form $(N + k', k')$, that is, all the codes that composite the code have exactly $N$ parity symbols, thus the number of parity symbols has to be some multiple of $N$.
\end{IEEEproof}

\begin{IEEEproof}[Proof of Corollary~\ref{corollary:maxsym}]
    This proof follows directly from Lemma~\ref{lemma:achiev} by noticing that:
    \begin{enumerate}
        \item The achievability can be written as
        \begin{align}
        	\bT^{(g)}[j] &\geq \frac{N n}{n - k}\left(1 - \sum_{\ell = 1}^{j-1} \frac{k^{(g)}[\ell]}{n} \right) - 1.
    	\end{align}
    	This is because we can always choose to recover symbols that could be recovered with a small delay later by introducing artificial delay.
    	\item Constraining $k^{(g)}[\ell] \leq k^{\textrm{con}}[\ell]$. 
    \end{enumerate}
\end{IEEEproof}

\begin{IEEEproof}[Proof of Lemma~\ref{lemma:achievablelowercorner}]
	Let us consider using $(T + 1 - N_3)c$ concatenations of a $\frac{T + 1 - N_3 - N_2}{T + 1 - N_2}$ interleaving MDS code in the second hop and $(T + 1 - N_2)c$ concatenations of a $\frac{T + 1 - N_3 - N_1}{T + 1 - N_3}$ code in the first link. Finally, let us fix $n = (T + 1 - N_2)(T + 1 - N_3)c$. Note that, under such conditions, we achieve the desired $R_1$ from the assumption of the Lemma. The constant $c$ is an auxiliary constant that should satisfy that the number of symbols in each timeslot is integer for all links. 
	
	Under such conditions, the delay spectrum of the second hop is uniform with $(T + 1 - N_3)c$ symbols from $N_3$ to $T - N_2$, and the delay spectrum of the first link is $(T + 1 - N_2)c$ symbols from $N_1$ to $T - N_3$. Therefore, we have the following constraint on the second link
	\begin{align}
		\mathbf{G}^{\textrm{con}} = \begin{bmatrix}
			(T - N_3, (N_2 - N_3)c)\\ (T - N_3 - 1, (N_2 - N_3)c)\\ \vdots \\ (N_1, (N_2 - N_3)c) \\ (N_1 + 1, (T + 1 - N_3)c) \\ \vdots \\ (N_2, (T + 1 - N_3)c)
		\end{bmatrix} .
	\end{align}

	Under this constraint, applying Corollary~\ref{corollary:maxsym}, we have the following conditions for achievability on the second link
	\begin{align}
		k_2 &\leq n - \frac{n N_2 }{T - N_3 + 1} \label{eq:single_user_bound}\\
		k_2 &\leq n - \frac{n N_2 (1 - \frac{(N_2 - N_3)c}{n}) }{(T - N_3 - 1) + 1}\nonumber\\
		k_2 &\leq n - \frac{n N_2 (1 - 2\frac{(N_2 - N_3)c}{n}) }{(T - N_3 - 2) + 1} \nonumber\\
		&\vdots \nonumber\\
		k_2 &\leq n - \frac{n N_2 (1 - (T + 1 - N_3 - N_1)\frac{(N_2 - N_3)c}{n}) }{N_1} \label{eq:delay_spectrum_constraint_bound} \\
		k_2 &\leq n - \frac{n N_2 (1 - (T + 1 - N_3 - N_1)\frac{(N_2 - N_3)c}{n} - \frac{(T + 1 - N_3)c}{n}) }{N_1 - 1} \nonumber\\
		k_2 &\leq n - \frac{n N_2 (1 - (T + 1 - N_3 - N_1)\frac{(N_2 - N_3)c}{n} - 2\frac{(T + 1 - N_3)c}{n}) }{N_1 - 2} \nonumber\\
		&\vdots \nonumber\\
		k_2 &\leq n - \frac{n N_2 (1 - (T + 1 - N_3 - N_1)\frac{(N_2 - N_3)c}{n} - (N_1 - N_2)\frac{(T + 1 - N_3)c}{n}) }{N_2} \label{eq:sum_rate_bound}
	\end{align}

	We note that equation \eqref{eq:single_user_bound} represents the single user upper bound, that is, $R_2 \leq C(T - N_3, N_2)$. Equation \eqref{eq:sum_rate_bound} represents the sumrate bound, i.e., $R_1 + R_2 \leq C(T - N_2, N_3)$. While there are many other constraints to be considered, we now show that they are dominated by \eqref{eq:delay_spectrum_constraint_bound} and \eqref{eq:sum_rate_bound} when $R_1 = C(T - N_3, N_1)$. In order to do that, let us first consider the function
	\begin{align}
		f(x) = n - n \frac{N_2 (1 - x \frac{(N_2 - N_3)c}{n})}{(T - N_3 - x) + 1}
	\end{align}
	with derivative
	\begin{equation}
		f'(x) = -n N_2 \left[ -\frac{\frac{(N_2 - N_3)c}{n}}{(T - N_3 - x) + 1}  +  \left(1 - x \frac{(N_2 - N_3)c}{n} \right) \frac{1}{(T + 1 - N_3 - x)^2} \right].
	\end{equation}
	It can be shown that, for $T \geq 2N_2 - N_3 - 1 = c_1$, this derivative is always negative. Further, recall that, in order for the second hop to be the bottleneck (i.e., the scenario we are interested on), we require $T \geq \frac{1}{2}\left( \sqrt{N_1^2 - 4N_3(N_2 - N_3)} + N_1 + 2N_2 - 2 \right) = c_2$. It can be shown that, for any $N_1 \geq N_2 \geq N_3$, we have $c_2 \geq c_1$, that is, in our regime of operation, this derivative is always negative. From this, it follows that \eqref{eq:delay_spectrum_constraint_bound} is the minimum among all constraints from \eqref{eq:single_user_bound} to \eqref{eq:delay_spectrum_constraint_bound}. 
	
	Similarly, we consider the function
	\begin{align}
		g(x) = n - n N_2 \frac{1 - (T + 1 - N_3 - N_1)\frac{(N_2 - N_3)c}{n} - x \frac{(T + 1 - N_3)c}{n}}{N_1 - x}
	\end{align}
	with derivative
	\begin{equation}
		g'(x) = -n N_2 \left[ - \frac{ \frac{(T + 1 - N_3)c}{n} }{N_1 - x}  + \frac{1 - (T + 1 - N_3 - N_1)\frac{(N_2 - N_3)c}{n} - x \frac{(T + 1 - N_3)c }{n} }{(N_1 - x)^2} \right].
	\end{equation}
	It can be shown that the derivative is negative if
	\begin{equation}
		(T + 1 - N_3)(T + 1 - N_2 - N_1) - (T + 1 - N_3 - N_1)(N_2 - N_3) \geq 0
	\end{equation}
	and positive otherwise. That is, either the derivative is always negative and \eqref{eq:sum_rate_bound} is the minimum, or the derivative is always positive and \eqref{eq:delay_spectrum_constraint_bound} is the minimum. Dividing both sides by $n$ (in order to obtain $R_2 = k_2/n$) and then substituting $n = (T + 1 - N_2)(T + 1 - N_3)c$ in both equations completes the proof.
\end{IEEEproof}

\begin{IEEEproof}[Proof of Corollary~\ref{corollary:minimumT}]
	This follows directly from the condition for $R_2' \geq C(T - N_2, N_3) - C(T - N_3, N_1)$ presented in the proof for Lemma~\ref{lemma:achievablelowercorner}.
\end{IEEEproof}

\begin{IEEEproof}[Proof of Lemma~\ref{lemma:intersectionpoint}]
	Consider the following scheme: in the second hop, we use $(T + 1 - N_3)c$ concatenations of a $\frac{T + 1 - N_3 - N_2}{T + 1 - N_2}$ code and the second link (of the first hop) uses $(T + 1 - N_3)c$ concatenations of a $\frac{T + 1 - N_2 - N_2}{T + 1 - N_2}$ code. Note that we have $n_2 = n_3 = (T + 1 - N_2)(T + 1 - N_3)c$. Then, we use $c(T + 1 - N_3)(N_2 - N_3)/(T + 1 - N_3 - N_1)$ codes of rate $\frac{T + 1 - N_1 - N_3}{T + 1 - N_3}$ in the first link. Again, $c$ is an auxiliary constant that should satisfy that the number of concatenations is integer, e.g., $c = (T + 1 - N_1 - N_3)$. Now, note that we have
	\begin{align}
		k_2 = c(T + 1 - N_2 - N_2)(T + 1 - N_3)\\
		k_1 = c(T + 1 - N_3) * (N_2 - N_3)\\
		k = c(T + 1 - N_3 - N_2)(T + 1 - N_3)
	\end{align}
	and we have exactly $k$ symbols being transmitted in the bottleneck. Furthermore, it is easy to see that the delay spectrum of these codes achieve the desired overall delay under $N_1$, $N_2$ and $N_3$ erasures. Then, it remains to find $n = \max(n_1, n_2)$. Note that if $n = n_2$, we achieve the sum capacity. Thus, the condition for achieving the sum capacity is
	\begin{align}
		c(T + 1 - N_2)(T + 1 - N_3 - N_1) \geq c(T + 1 - N_3)(N_2 - N_3)
	\end{align}
	
	Note that this condition is equivalent to the condition that we achieve the sum capacity at some point (from Corollary~\ref{corollary:minimumT}). Further, note that, if this condition holds, we have
	\begin{align}
		R_1 &= \frac{c(T + 1 - N_3) * (N_2 - N_3)}{ c(T + 1 - N_2)(T + 1 - N_3) }\\
		&= \frac{N_2 - N_3}{T + 1 - N_2}
	\end{align}
	which completes the proof.
\end{IEEEproof}
\begin{remark}
	Note that, if the condition for achievability of the sum capacity holds, we have
	\begin{align}
		\frac{N_2 - N_3}{T + 1 - N_2} \leq C(T - N_3, N_1)
	\end{align}
	and
	\begin{align}
		C(T - N_2, N_2) \geq C(T - N_2, N_3) - C(T - N_3, N_1).
	\end{align}
	That is, the point that achieves the sum capacity in Lemma~\ref{lemma:intersectionpoint} is to the ``left'' and ``up'' in the rate region than the point that achieves it in Lemma~\ref{lemma:achievablelowercorner}. In fact, this is the point with highest $R_2$ that can achieve the sum capacity using the family of schemes we propose.
\end{remark}

\begin{IEEEproof}[Proof of Theorem~\ref{theorem:R2givenR1}]
    For simplicity, consider the code used to achieve $R_1$ in the proof of Lemma~\ref{lemma:achievablelowercorner}, and consider that we delete information symbols uniformly across all timeslots in order to achieve some desired $R_1$ (while maintaining $n = (T + 1 - N_3)(T + 1 - N_2)c$). In that case, in the first link, we have $\frac{R_1 n}{T + 1 - N_1 - N_3}$ symbols per timeslot in the first link. We denote by $R_{bn} = C(T - N_3, N_2) = \frac{T + 1 - N_3 - N_2}{T + 1 - N_2}$. In this case, we have that the constraints change to
    \begin{align}
    	k_2 &\leq n - \frac{n N_2 }{T - N_3 + 1} \label{eq:generalR1first} \\
    	k_2 &\leq n - \frac{n N_2 (1 - R_{bn} \frac{1}{T + 1 - N_2 - N_3} + R1 \frac{1}{T + 1 - N_1 - N_3})}{(T - N_3 - 1) + 1}\nonumber\\
    	k_2 &\leq n - \frac{n N_2 (1 - R_{bn} \frac{2}{T + 1 - N_2 - N_3} + R1 \frac{2}{T + 1 - N_1 - N_3})}{(T - N_3 - 2) + 1} \nonumber\\
    	&\vdots \nonumber\\
    	k_2 &\leq n - \frac{n N_2 (1 - R_{bn} \frac{T + 1 - N_1 - N_3}{T + 1 - N_2 - N_3} + R1 \frac{T + 1 - N_1 - N_3}{T + 1 - N_1 - N_3})}{N_1} \label{eq:generalR1second}  \\
    	k_2 &\leq n - \frac{n N_2 (1 - R_{bn} \frac{T + 1 - N_1 - N_3}{T + 1 - N_2 - N_3} + R1 \frac{T + 1 - N_1 - N_3}{T + 1 - N_1 - N_3}) - \frac{(T + 1 - N_3)c }{n} }{N_1 - 1} \nonumber\\
    	k_2 &\leq n - \frac{n N_2 (1 - R_{bn} \frac{T + 1 - N_1 - N_3}{T + 1 - N_2 - N_3} + R1 \frac{T + 1 - N_1 - N_3}{T + 1 - N_1 - N_3}) - 2\frac{(T + 1 - N_3)c }{n} }{N_1 - 2} \nonumber\\
    	&\vdots \nonumber\\
    	k_2 &\leq n - \frac{n N_2 (1 - R_{bn} \frac{T + 1 - N_1 - N_3}{T + 1 - N_2 - N_3} + R1 \frac{T + 1 - N_1 - N_3}{T + 1 - N_1 - N_3}) - (N_1 - N_2)\frac{(T + 1 - N_3)c }{n} }{N_2} \label{eq:generalR1third}
    \end{align}
    
    Similar to before, let us conider the function
	\begin{align}
		f(x) = n - n N_2 \frac{1 - R_{bn}\frac{x}{T + 1 - N_2 - N_3} + R_1 \frac{x}{T + 1 - N_1 - N_3} }{(T - N_3 - x + 1)}
	\end{align}
	with derivative
	\begin{equation}
		f'(x) = -n N_2 \left[\frac{ \frac{R_1}{T + 1 - N_1 - N_3} - \frac{R_{bn}}{T + 1 - N_2 - N_3}}{T + 1 - N_3 - x}  + \frac{\left(1 - \frac{xR_{bn}}{T + 1 - N_2 - N_3} +  \frac{xR_1}{T + 1 - N_1 - N_3}   \right)}{(T + 1 - N_3 - x)^2} \right].
	\end{equation}
	Then, note that
	\begin{align}
		&f'(x) \leq 0 \Rightarrow \nonumber \\
		& \left[\left(\frac{R_1(T + 1 - N_3)}{T + 1 - N_1 - N_3} - \frac{(T + 1 - N_3)}{T + 1 - N_2}\right)  + 1 \right] \geq 0.
	\end{align}
	That is, for any set of parameters and $R_1$, this derivative is always positive or always negative, thus, the minimum of all expressions between \eqref{eq:generalR1first} and \eqref{eq:generalR1second} is always in the extremes, i.e., one of these two expressions.
	
	Similarly, let us consider the function
	\begin{align}
		& g(x) = \nonumber \\
		& n - \frac{n N_2 (1 - R_{bn} \frac{T + 1 - N_1 - N_3}{T + 1 - N_2 - N_3} + R1 \frac{T + 1 - N_1 - N_3}{T + 1 - N_1 - N_3} - x\frac{(T + 1 - N_3)c }{n}) }{N_1 - x}
	\end{align}
	with derivative
	\begin{multline}
		g'(x) = -n N_2 \left[ -\frac{\frac{(T + 1 - N_3)c}{n}}{N_1 - x}  \right. \\+ \left. \frac{(1 - R_{bn} \frac{T + 1 - N_1 - N_3}{T + 1 - N_2 - N_3} + R1 \frac{T + 1 - N_1 - N_3}{T + 1 - N_1 - N_3} - x\frac{(T + 1 - N_3)c }{n})}{(N_1 - x)^2}   \right].
	\end{multline}
	And again, note that
	\begin{align}
		&g'(x) \leq 0 \Rightarrow \nonumber \\
		&\left[ 1 - \frac{T + 1 - 2N_1 - N_3}{T + 1 - N_2} + R1 \frac{T + 1 - N_1 - N_3}{T + 1 - N_1 - N_3}   \right] \geq 0.
	\end{align}
	That is, again, the derivative is always negative or always positive depending on the set of parameters and $R_1$, which means the minimum of all expressions between \eqref{eq:generalR1second} and \eqref{eq:generalR1third} lies in the extreme points, i.e., one of these two expressions. Therefore, it suffices to take only these three points into consideration. Finding $R_2 = k_2/n$ and subtituting $n = (T + 1 - N_3)(T + 1 - N_2)c$ completes the proof.
\end{IEEEproof}

\begin{IEEEproof}[Proof of Lemma~\ref{lemma:timesharing}]
    The proof follows directly from concatenating the two streaming codes used. More specifically, the first user concatenates $A$ copies of the $(n_1, k_1)$ code with $B$ copies of the $(n_1', k_1')$ code, the second users concatenates $A$ copies of the $(n_2, k_2)$ code with $B$ copies of the $(n_2', k_2')$ code, and the relay concatenates $A$ copies of the $(n_3, k_1 + k_2)$ code with $B$ copies of the $(n_3', k_1' + k_2')$ code. Then, from the fact that these codes are able to achieve delay $T$ under $(N_1, N_2, N_3)$ erasures, it follows from Lemma~\ref{lemma:conc} that the concatenation of these codes is also able to achieve delay $T$ under $(N_1, N_2, N_3)$ erasures.
\end{IEEEproof}

\begin{IEEEproof}[Proof of Lemma~\ref{lemma:uppercorner}]
    This follows directly from applying Lemma~\ref{lemma:timesharing} to the streaming codes described for the CSWDF, that is
    \begin{align}
        &(T + 1 - N_3, 0, T + 1 - N_1, T + 1 - N_1 - N_3, 0)_{\mathbb{F}} \\
        &(0, T + 1 - N_3, T + 1 - N_2, 0, T + 1 - N_2 - N_3)_{\mathbb{F}}
    \end{align}
    Specifically, we use $A = N_2 - N_3$ and $B = T + 1 - N_1$, therefore, we have a streaming code certain to achieve the required delay, with parameters
    \begin{align}
        k_1 &= (N_2 - N_3)(T + 1 - N_1 - N_3) \\
        n_1 &= (N_2 - N_3)(T + 1 - N_3) \\
        k_2 &= (T + 1 - N_1)(T + 1 - N_2 - N_3)\\
        n_2 &= (T + 1 - N_1)(T + 1 - N_3) \\
        n_3 &= (N_2 - N_3)(T + 1 - N_1) + (T + 1 - N_1)(T + 1 - N_2).
    \end{align}
    Then, it remains to find $n$ and the respective rates. Note that
    \begin{align}
        n_3 = (T + 1 - N_1)(T + 1 - N_3) = n_2.
    \end{align}
    Finally, we need to show that
    \begin{align}
        T + 1 - N_1 \geq N_2 - N_3
    \end{align}
    which implies $n_3 \geq n_1$. The proof is as follows: recall that we assume $$T \geq \frac{1}{2}\left( \sqrt{N_1^2 - 4N_3(N_2 - N_3)} + N_1 + 2N_2 - 2 \right)$$ thus, it suffices to show that this assumption implies that the above condition holds. We start by computing
    \begin{align}
        &\frac{1}{2}\left( \sqrt{N_1^2 - 4N_3(N_2 - N_3)} + N_1 + 2N_2 - 2 \right) \nonumber  - (N_1 + N_2 - N_3 - 1) \\
        = &\frac{1}{2}\left( \sqrt{N_1^2 - 4N_3(N_2 - N_3)} - N_1 + 2 N_3 \right).
    \end{align}
    In order to have $T + 1 - N_1 \geq N_2 - N_3$ guaranteed, we need this expression to be greater than zero, or
    \begin{align}
        \left( \sqrt{N_1^2 - 4N_3(N_2 - N_3)}  \right) \geq N_1 - 2 N_3.
    \end{align}
    Now, note that, if $N_1 - 2N_3 \leq 0$, then this condition holds. If $N_1 - 2N_3 \geq 0$, then we can rewrite the condition as
    \begin{align}
        N_1^2 - 4N_3(N_2 - N_3) &\geq N_1^2 - 4N_1 N_3 + 4 N_3^2 \\
        - 4N_3 N_2 &\geq -4 N_1 N_3 \\
        N_1 N_3 &\geq N_2 N_3 \\ 
        N_1 &\geq N_2.
    \end{align}
    Finally, recall that this is also an assumption, therefore, in this regime of operation, we always have $n = n_3 = n_2 \geq n_1$. Finally, we can compute the rates
    \begin{align}
        R_1 &= \frac{(N_2 - N_3)(T + 1 - N_1 - N_3) }{ (T + 1 - N_1)(T + 1 - N_3)} \\
        R_2 &= \frac{(T + 1 - N_1)(T + 1 - N_2 - N_3)}{(T + 1 - N_1)(T + 1 - N_3)}\nonumber\\ &= \frac{T + 1 - N_2 - N_3}{T + 1 - N_3} = C(T - N_3, N_2)
    \end{align}
    which are the rates described in the lemma.
\end{IEEEproof}

\begin{IEEEproof}[More details on Proof of Lemma~\ref{lemma:weaksr}]
    This follows directly from the fact that, under such condition, we can use the same scheme used for the strong source-relay bottleneck, that is, a simple CSWDF with one concatenation of each single-user capacity-achieving code (i.e. $A = B = 1$). In that case, we have, as before, $n_3 = (T + 1 - N_1) + (T + 1 - N_2)$, and $n_1 = n_2 = T + 1 - N_3$. Then, if $n_1 \geq n_3$, we achieve the capacity, since the rates will be $R_1 = k_1/n_1 = C(T - N_3, N_1)$ and $R_2 = k_2/n_2 = C(T - N_3, N_2)$. To complete the proof, it is sufficient to notice that the condition in the Lemma is exactly the condition for $n_1 \geq n_3$. 
\end{IEEEproof}

\bibliographystyle{IEEEtran}
\bibliography{eladd.bib}

\end{document}